\documentclass[preprint]{aastex}
\usepackage{psfig}
\usepackage{lscape}

\def\sax{{\sl BeppoSAX }}
\def\rxte{{\sl RXTE}}

\def\exo{{\sl EXOSAT }}

\def\einstein{{\sl Einstein }}

\def\xte{{\sl RXTE }}

\def\xmm{{\sl XMM-Newton }}
\def\chandra{{\sl Chandra }}

\def\herx1{Her~X-1}
\def\ergsec{\hbox{erg s$^{-1}$ }}

\def\fexxv{Fe~{\sc xxv}}
\def\fexxvi{Fe~{\sc xxvi}}
\def\fexvii{Fe~{\sc xvii}}
\def\fei{Fe~{\sc i}}

\def\fex{Fe~{\sc x}}
\def\nvi{N~{\sc vi}}
\def\nvii{N~{\sc vii}}

\def\neix{Ne~{\sc ix}}
\def\oviii{O~{\sc viii}}
\def\ovii{O~{\sc vii}}
\def\ovi{O~{\sc vi}}
\def\mgxi{Mg~{\sc xi}}

\def\sixiii{Si~{\sc xiii}}
\def\sxvi{S~{\sc xvi}}

\def\it{\sl}

\def\lapp{\ifmmode\stackrel{<}{_{\sim}}\else$\stackrel{<}{_{\sim}}$\fi}

\def\gapp{\ifmmode\stackrel{>}{_{\sim}}\else$\stackrel{>}{_{\sim}}$\fi}
\def\spose#1{\hbox to 0pt{#1\hss}}

\def\approxlt{\mathrel{\spose{\lower 3pt\hbox{$\sim$}}
        \raise 2.0pt\hbox{$<$}}}
\def\approxgt{\mathrel{\spose{\lower 3pt\hbox{$\sim$}}
        \raise 2.0pt\hbox{$>$}}}

\slugcomment{Submitted for publication to The Astrophysical Journal}

\shorttitle{PHOTOIONIZATION IN HER~X-1}

\shortauthors{Ji et al.}

\begin{document}

\title{The Photoionized Accretion Disk in Her X-1}

\author{L. Ji$^{1}$, N. Schulz$^{1}$, M. Nowak$^{1}$, H.L. Marshall$^{1}$, \& T. Kallman$^{2}$}
\affil{$^{1}$MIT Kavli Institute for Astrophysics and Space Research, Cambridge, MA ji@space.mit.edu; }
\affil{$^{2}$NASA Goddard Space Flight Center, NASA, Greenbelt, MD}

\begin{abstract}
We present an analysis of several high-resolution Chandra grating
observations of the X-ray binary pulsar Her X-1. With a total exposure
of 170 ks, the observations are separated by years and cover three
combinations of orbital and super-orbital phases. Our goal is to
determine distinct properties of the photoionized emission and its
dependence on phase-dependent variations of the continuum.  We find
that the continua can be described by a partial covering model which
above 2 keV is consistent with recent results from \rxte~ studies and
at low energies is consistent with recent \xmm and \sax studies.  Besides a powerlaw
with fixed index, an additional thermal blackbody of 114 eV is
required to fit wavelengths above 12 \AA~ ($\sim$ 1 keV). We find that
likely all the variability is caused by highly variable absorption
columns in the range (1 -- 3)$\times 10^{23}$ cm$^{-2}$. Strong Fe K line
fluorescence in almost all observations reveals that dense, cool
material is present not only in the outer regions of the disk but
interspersed throughout the disk.  Most spectra show strong line
emission stemming from a photoionized accretion disk corona. We model
the line emission with generic thermal plasma models as well as with
the photoionization code XSTAR and investigate changes of the
ionization balance with orbital and superorbital phases. Most
accretion disk coronal properties such as disk radii, temperatures,
and plasma densities are consistent with previous findings for the low
state.  We find that these properties change negligibly with respect to
orbital and super-orbital phases. A couple of the higher energy lines
exhibit emissivities that are significantly in excess of expectations
from a static accretion disk corona. 


\end{abstract}

\section{Introduction}
The mechanisms involved in accretion processes of X-ray binaries are
still quite poorly understood specifically when it comes to feedback
effects and how they affect the overall X-ray luminosity. In low-mass
X-ray binaries (LMXBs) a neutron star or a stellar black hole accretes
matter from its companion entirely via Roche-lobe overflow and a
steady accretion disk is formed. The bulk of the gravitational energy
is released in the form of X-rays near the compact object at
luminosities of up to over $10^{38}$ \ergsec. LMXBs have been
classified into two major categories based on their correlated
spectral and timing behavior into Z- and atoll sources, where the
former predominantly radiate near their Eddington luminosities, the
latter radiate generally an order of magnitude
lower~\citep{hasinger89, schulz89}.

Although \herx1 \citep{tananbaum72} is not a typical LMXB, its bulk
mass accretion also comes from Roche-lobe overflow and it possesses a
sizeable accretion disk. With respect to the Atoll and Z-source
classification scheme, \herx1 has been associated with neither.
However, properties of accretion disk coronae (ADCs), as observed in
some Atoll sources viewed at high inclination angles such as 4U
1822-37~\citep{cottam01}, 2S 0921-63~\citep{kallman03}, or EXO 0748-767
\citep{mario03}, have also been observed in \herx1
\citep{jimenez2005}. From the point of accretion dynamics (i.e. the structures
of the outer accretion disk), we expect \herx1 to behave
very much like these Atoll sources.  There are two primary differences
with respect to LMXBs in the \herx1 system besides the larger
companion mass of $2.3\pm 0.3 ~\rm M_{\odot}$. One is a high magnetic
field of the accreting neutron star leading to the 1.24 s X-ray
pulsation period and therefore some fraction of its X-rays come from
the polar cap accretion columns. LMXBs have lower magnetic fields and
most of them likely lack this component in their emissions although 4U1822-37
has a detected X-ray pulse with a period of 0.59 s \citep{jonker01}. 
\herx1, besides its
orbital period of P$_{orb}$ = 1.7 days, also possesses a $P_{\Psi}\sim
35$ day superorbital period likely due to some form of disk warp.
Such periods have also been observed in systems with high-mass
companions (HMXBs) such as LMC X-4 and SMC X-1. Evidence that a
stellar wind is affecting the observability of disk signatures in the
ultra-violet has been presented recently~\citep{boro07}.  Either
accretion disk winds~\citep{schandl94} or radiation pressure
effects~\citep{maloney97} are considered to cause this precessing
warp.

Emissions from accretion in X-ray binaries are observed from about
several eV up to hundreds of keV. Each energy band carries information
about specific emission processes, the nature of the radiative source
itself as well as its location. In rough terms it goes like the
following scheme. Optical emission is usually associated with the
companion star HZ Her which is indirectly affected by heating from the
X-ray source~\citep{middleditch76, still97}. The latter primarily
manifests itself in the UV which also hosts the bulk of the continuum
emission from the accretion disk. UV emission lines, the strongest
include \ion{C}{4}, \ion{N}{5}, \ion{Si}{4}, result from
photoionization of large portions of the accretion
disk~\citep{howard1983,cheng1995,raymond1993,ko1994}.  In his classic
and comprehensive model of an X-ray-illuminated surface of the
accretion disk in LMXBs, \citet{raymond1993} produced UV lines
consistent with observations in LMXBs and \herx1 as well as predicted
soft X-ray line emission on top of the blackbody emission from 
the inner disk \citep{oosterbroek01}.  Most of the hard X-rays above 10 -- 20 keV
come from the central source in form of Compton scattering of thermal
emission from the central source as well as synchrotron emission from
the accretion column. The discovery of a 42 keV synchrotron line
allowed a magnetic field determination of $3.5\times 10^{12} $G
\citep{truemper78, fiume98}. A recent account of X-ray emissions
observed with \xte can be found in ~\citet{kuster05}.

In this paper we concentrate on X-ray properties in the 0.1 to 8 keV
band with emphasis on orbital and super-orbital influence on
photoionized properties of the accretion disk.  \sax observations
obtained during the main-on state described a spectral continuum
consisting of a powerlaw of index 0.74 and a strong 0.093 keV
blackbody component plus a broad Fe L line complex at 0.937 keV and an
Fe K line complex at 6.4 keV ~\citep{oosterbroek01}. More recently
multiple \xte observations scanned the turn-on, a transition from the
low state to the main-on state in the 35 day cycle, and found an
evolution in the spectral continuum dominated by scattering and
partial covering absorption~\citep{kuster05}.  In a first high
resolution pilot study, \citet{jimenez2005} identified an extended
accretion disk corona in \herx1's low state using high resolution
X-ray spectra from \chandra. More than two dozen recombination lines
from \fexxvi~ (1.78~\AA) to \nvi~ (29.08~\AA) were detected and
emission radii between $8\times10^{10}$ and $1\times10^{11}$ cm were
derived. Line optical depth diagnostics were consistent with a
flattened atmosphere. These properties were derived for the low state
and the question remains how they change with respect to orbital and
super-orbital phase. The pilot study also deduced an overabundance of
nitrogen relative to the other metals, which had been predicted from
UV line studies.

The following analysis covers half a dozen exposures of \herx1 with
the High Energy Transition Grating Spectrometer (HETGS) onboard
\chandra~\citep{canizares05} including the one presented by
\citet{jimenez2005}.  In \S2 we present the observations, light curves
and continuum fits. \S3 to \S5 cover line detections, emission
measures, and the discussion of their implications, respectively.

\section{Chandra Observations}

The \chandra archive to date includes total of six HETGS observations of
\herx1 that were performed in the timed event (TE) mode. The basic
observation statuses are summarized in Table \ref{tab-obsid}. These observations span
several different phase combinations with respect to orbital and
super-orbital phase, the low state (OBSID 2749), the turn-on (OBSIDs
3821, 3822), the main-on (OBSIDs 4585, 6149), and the turn-off (OBSID
6150). We reprocess all the observations using CIAO Version 4.1 with
the most recent CALDB products. We restrict the analysis to
observations obtained in TE mode to maintain the most coherent
calibration status.  Some observations in the archive were taken in
continuous clocking mode during a specifically bright episode of the
source. Here we opt to wait for better calibration in order to relate
the CC mode data to the TE mode data.

For each observation, we then re-determine the positions of
the 0th order centroid for optimal wavelength accuracy.
We extract all 1st order spectra in HEG and MEG and combine them.
Light curves and spectral analysis are based on ISIS Version 1.4.7 
\citep{houck00}. 

\subsection{Light Curve Analysis}

\begin{figure}
\centering
\mbox{
\includegraphics[width=0.5\textwidth]{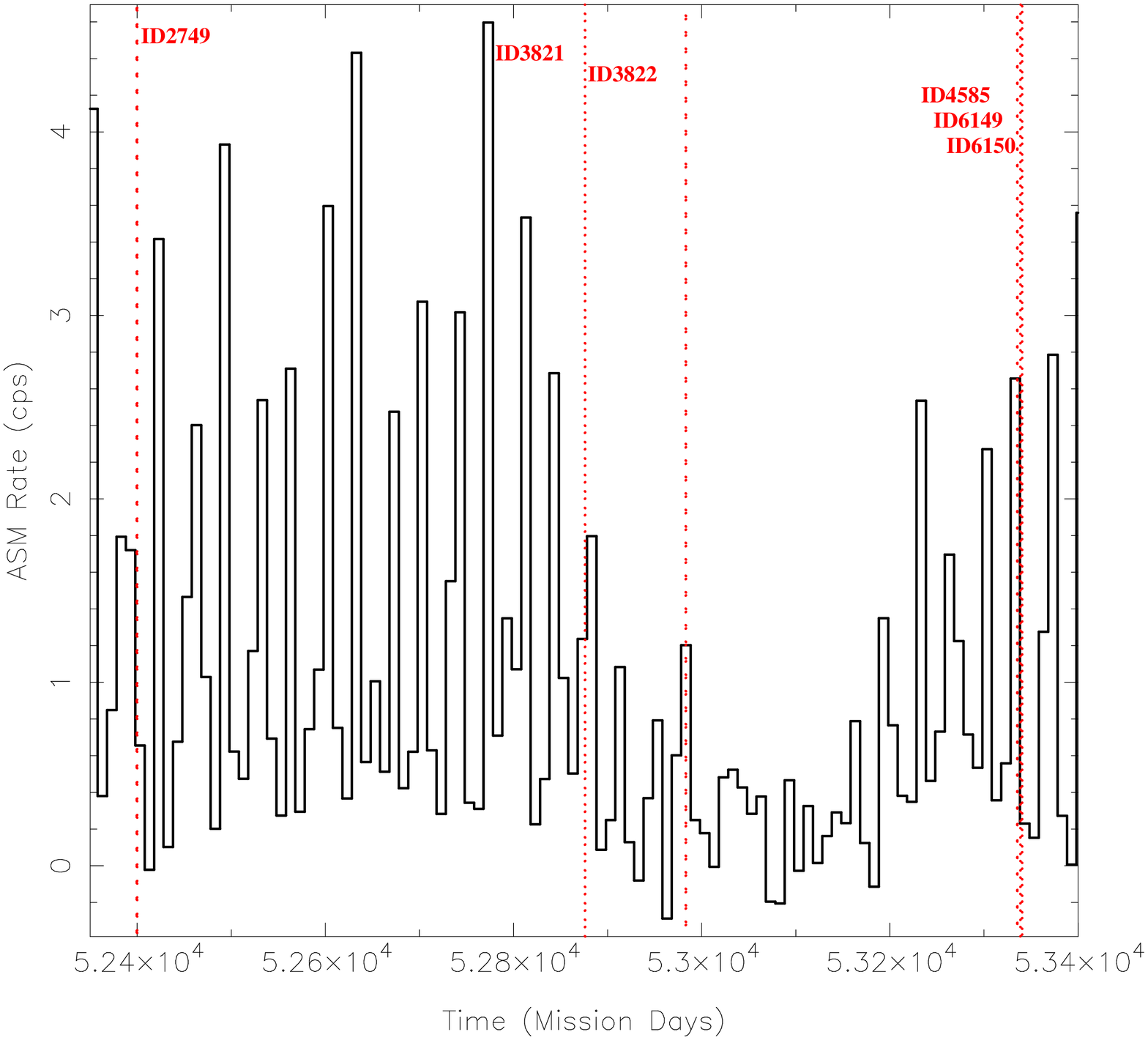} 
\includegraphics[width=0.48\textwidth]{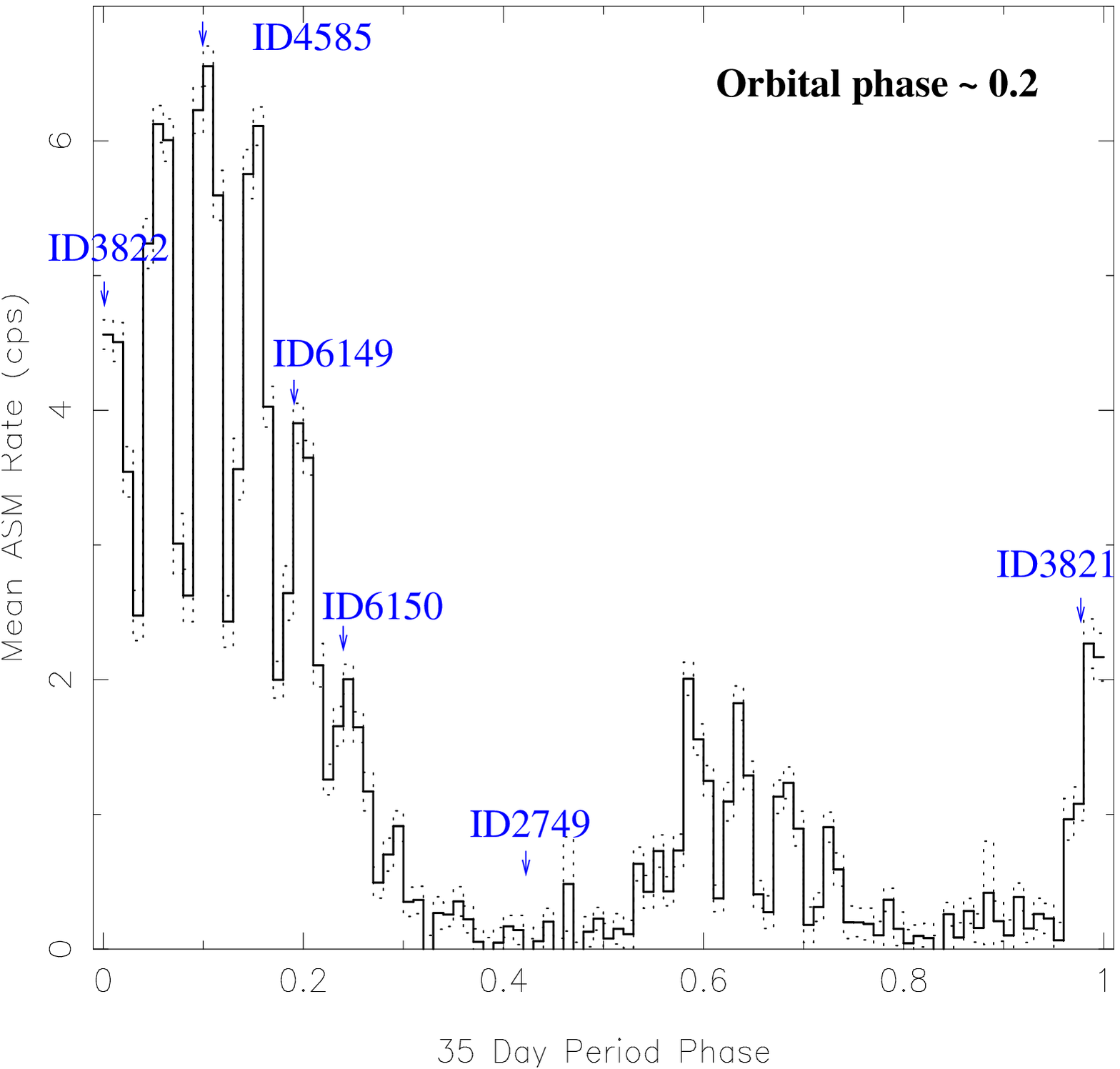}
}
\caption{Left: Her X-1 ASM light curve; Right: averaged X-ray light 
         curves of Her X-1 corresponding to cycles with
         turn-ons near orbital phase $\sim~0.2$. 
	 }
\label{fig-asmlc}
\end{figure}

In order to generate a long-term average X-ray light curve we use all
the observations obtained from the All-Sky Monitor (ASM) instrument on
board the {\it Rossi X-Ray Timing Explorer (RXTE)}. The left panel in
Figure~\ref{fig-asmlc} shows this lightcurve and highlights the phase
locations for all \chandra observations from Table 1. We show the
light curve for two orbital reference phases.  We superpose all 35 day
light curves corresponding to cycles with turn-ons near orbital phase
of about 0.2 s (right panel, similar for 0.7 s orbital phase).  The 35
day phase ephemeris was provided by R. Staubert through private
contact \citep{staubert07}.

\begin{figure}[h]
  \includegraphics[width=1.0\textwidth]{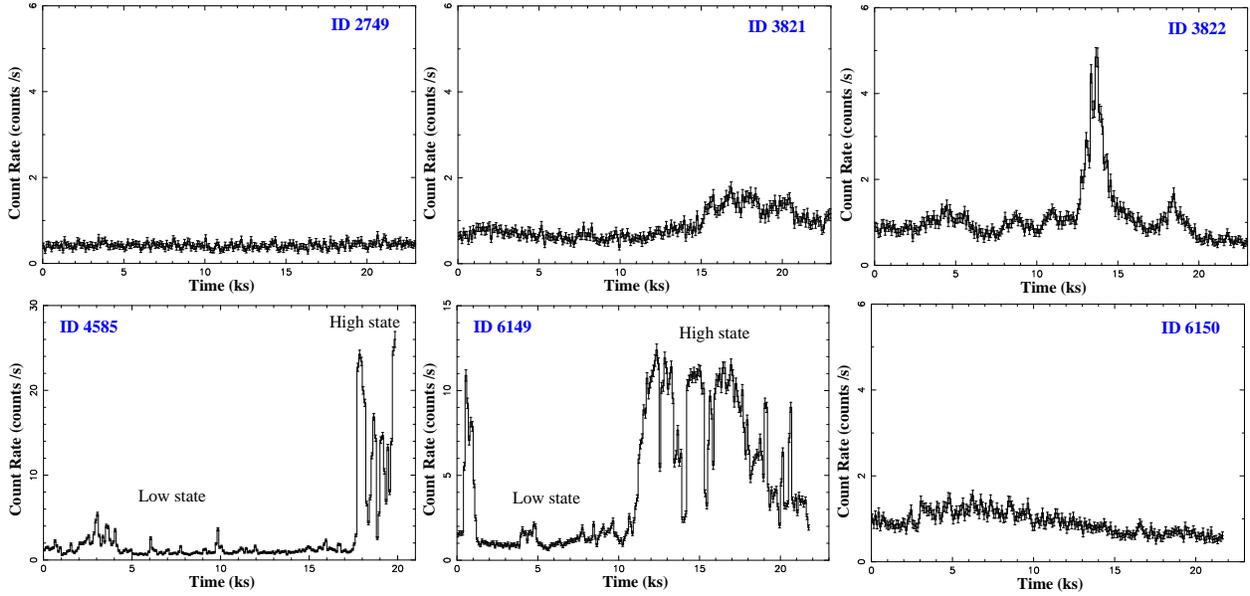}
 \caption{Light curve for all phases with $\lambda > 1.5 ~\AA $}
 \label{fig-lc1.5}
\end{figure}

The HETGS light curves were integrated in bins of 100s and for
wavelengths with $\lambda > 1.5 ~\AA $ (Figure \ref{fig-lc1.5}).
Count rates vary from about $0.2 \rm ~ct~ s^{-1}$ to $30\rm~ ct
~s^{-1}$.  OBSID 2749 is  in the off state, OBSID 6150 is close
to the off state and flux variations are small in both. There are some
small flares in the turn-on state (OBSIDs 3821 and 3822). Dramatic
flux variations by almost two orders of magnitude occur in the main-on
observations in OBSIDs 4585 and 6149.

\begin{figure}[h]
\mbox{
    \includegraphics[width=1.00\textwidth]{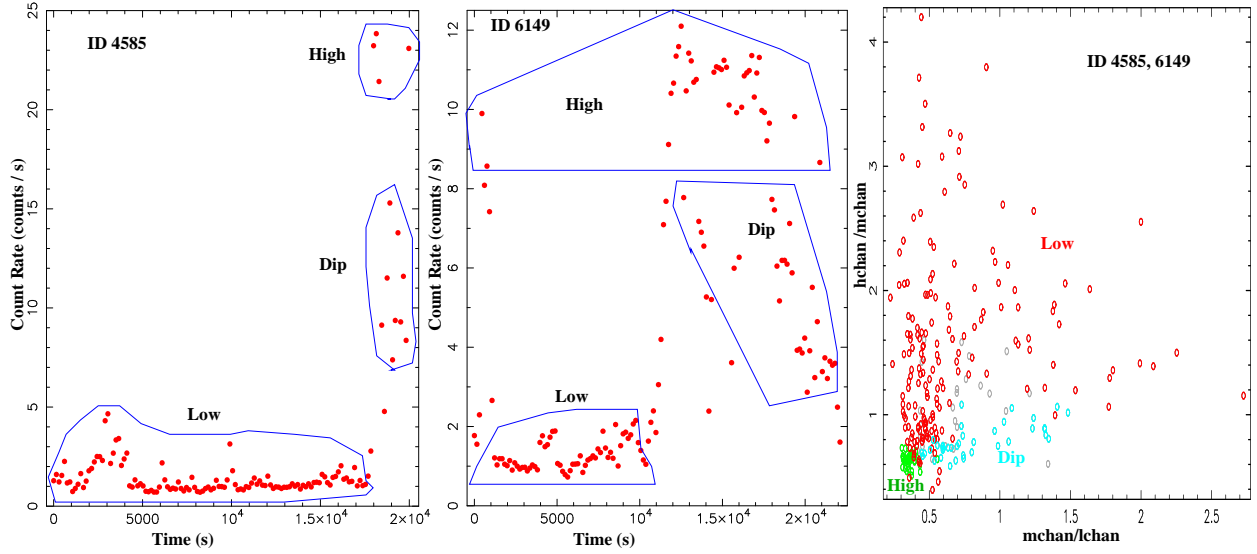}}
    \caption{Light curves for ID4585(left), ID6149(middle): separated into three regions(High, Low \& Dip). Right panel: color-color diagram for ID4585,6149, High(green), Low(red), Dip(cyan). Note: lchan:[5,25]\AA, mchan:[2.8,5.0]\AA, hchan:[1.0,2.8]\AA. See text for details. }
    \label{fig-main_sep}
\end{figure}

Significant flux variations indicate the existence of various
sub-states within single observations.  We separate the two
observations during main-on into high and low flux sub-states as well
as recognized dipping phases.  For the dipping phase during the
main-on, we further examine the color-color diagrams and separate
main-high, main-dip, and main-low phases. The left and middle panels of
Figure \ref{fig-main_sep} show the separated regions marked in the
light curve of OBSIDs 4585 and 6149 respectively.  The right panel of
Figure \ref{fig-main_sep} illustrates their positions in the
color-color diagram for the main-on stage. We compute the count rates
from three wavelength bands: a low band [5, 25] \AA, a middle band
[2.8, 5.0] \AA, and a high band [1.0, 2.8] \AA ~for each light curve
and calculate count rate ratios. The soft
color is the ratio of the middle to the high wavelength band,
the hard color ratio is the ratio of the low to the middle
wavelength band.

The main-high sub-state exhibits the lowest soft and hard ratios, with
values accumulating in the lower left corner of the color-color
diagram. The main-dip remains at low hard ratios, but expands to
higher soft ratios. In contrast, the main-low sub-state remains mostly
at low soft ratios but expands to higher hard ratios. In that respect,
dip and low states during the main-on phase appear to be qualitatively
different in their spectral evolution.  Thus although the color-color
diagram on the whole appears not very well structured, mostly due to a
somewhat erratic behavior during the low states, these spectral
variations do not appear atypical for atoll sources.

\subsection{Continuum Analysis}

Generally continuum modeling involves two components: absorbers and emitters.
Recent studies established powerlaw and blackbody functions to model the continuum in the \herx1 
\citep[see][]{oosterbroek01}.
Besides low foreground neutral absorption due to the interstellar medium, additional
neutral  and ionized absorbers have been applied
to fit the continuum during the dips \citep{trigo06}. Other authors favor a partial covering model
~\citep{brandt1996, kuster05}. Spectral changes in the different phases of 
\herx1 seem to require at least the latter~\citep{kuster05}.

Following that analysis we  model the spectrum of the emitting 
source with a powerlaw and a high energy exponential cutoff. 
Based on those {\it RXTE} observations of the turn-on phase 
of a 35 day cycle of \herx1, this continuum plus a partial 
covering model can well describe the high energy spectra 
in the [3, 17] keV range.
To be consistent with the \xte studies, we fix the powerlaw
photon index at 1.068 and the high energy cut-off at 21.5 keV in all fits.

\begin{figure}[h]
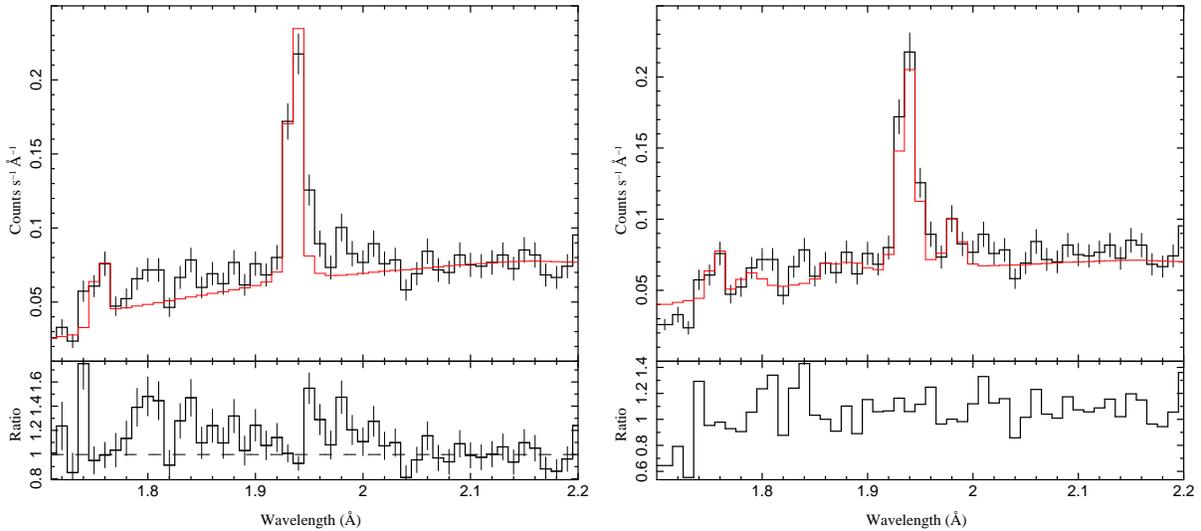

\centering
\mbox{
  \includegraphics[height=0.48\textwidth,angle=270]{./f4_a.ps}
  \includegraphics[height=0.48\textwidth,angle=270]{./f4_b.ps}}
\caption{ID3822: with(left) and without(right) local absorption.}
\label{fig-con3822}
\end{figure}

Located at high Galactic latitude, \herx1 suffers from very small reddening.
Photoelectric absorption from the line-of-sight ISM is fixed at
$N_{H}=1.3 \times 10^{18} ~\rm cm^{-2}$ (Jimenez-Garate et al. 2005).
We detected significant optical depth at the Fe K edges in all the observations indicating 
that significant amounts of absorption exist locally. A partial covering fraction absorption 
function matches well to these Fe K edges which is illustrated in Figure \ref{fig-con3822}.

\begin{figure}[h]
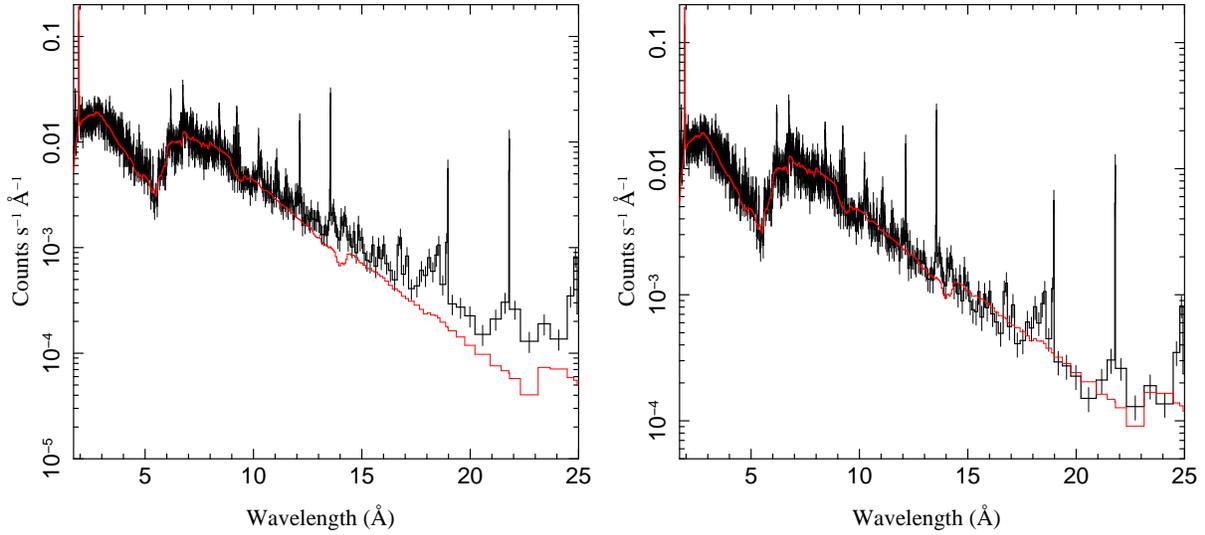

\mbox{
  \includegraphics[height=0.48\textwidth,angle=270]{./f5_a.ps}
  \includegraphics[height=0.48\textwidth,angle=270]{./f5_b.ps}}
\caption{ ID2749: with(Right) and without(left) blackbody component.}
\label{fig-con2749}
\end{figure}

All HETG fits also showed significant residuals above 12 \AA~ (below 1 keV)
requiring an additional component.
This is illustrated in Figure \ref{fig-con2749}. A 
thermal blackbody of 114 eV consistently fit the residuals in all cases. 

The best fit results for all phases are summarized in Table
\ref{tab-con}.  All phases exhibit substantial local
absorption on the order of $10^{23}~\rm cm^{-2}$ at large covering
fractions, which indicates a large fraction of local cold material
blocking our line of sight, even during the main-on maximum state
OBSID 4585.  In addition, we find that the higher the column density,
the higher the partial covering factors are. The column density of the
low, on or off stage is more than one magnitude larger than the one in
the high or dipper stage. This is consistent with the super-orbital
phase since main high (or dipper) stage suffers less blocking from the
central X-ray emitter or the local absorbers, which is confirmed by a
smaller covering factor and lower column density.
Due to the high inclination of the system ($i> 80$),
such local absorption is likely due to the accretion disk.

Table \ref{tab-con1} presents the best continuum fitting results
according to the different stages and sub-stages. The implications are
the same as those for Table \ref{tab-con}. Therefore, for the lines
from \S3 to \S5, we make our analysis based on the continuum presented
in Table \ref{tab-con}. We also apply the ``tbnew'' 
model \footnotemark \citep{wilms2000} which has
updated absorption cross-sections to replace the phabs model,
resulting in roughly twice the values of the fitted $N_{H}$.

\footnotetext {\tt {http://pulsar.sternwarte.uni-erlangen.de/wilms/research/tbabs/}}

The size of the thermal region can be derived from the flux of the
thermal blackbody component according to $L=\sigma T^{4} 4\pi R^{2}$:
$R(\rm km)=0.18361\sqrt{K(10^{-5})}/T^{2}(\rm keV) $, where $K$ is the
normalization of the blackbody in units of $10^{34} \rm~ ergs~
s^{-1}/(10 \rm kpc)^{2} $.  The size of the thermal component then
varies from about $10 \rm ~km$ to $100\rm ~km$, indicating the
emission does not originate from the neutron star surface but likely 
favors an inner disk origin.  Given a distance of 6.6 kpc,
\citet{fiume98} measured a flux with \sax in the [0.1, 200] keV band of
$7.2\times 10^{-9}\rm~ ergs~ s^{-1} cm^{-2}$, corresponding to a total
luminosity of about $3.8\times 10^{37} \rm~ ergs~ s^{-1}$. Their
observation was near the maximum of the 35 day X-ray intensity
cycle. The luminosity of our detection for the main-on maximum state
in OBSID 4585 is about $3.6\times 10^{37} \rm~ ergs~ s^{-1}$, which
is consistent with the \sax result.

Table \ref{tab-con} shows that luminosities in the [0.3, 10] keV band
for all phases are within a factor of a few except the one in low
state of OBSID 2749, which is almost one magnitude lower.  Given
stable accretion in \herx1, the luminosity of the central source ought
to be constant.  How real these fluctuations are needs to be
determined. It is interesting to note that these variations seem to
correlate with observed flux and partial covering absorption,
indicating that our model parameters for the partial coverage may
provide an incomplete description. The same correlation may also mean
that we see real variations in the heating of the ADC at different
phases.

\section{Fluorescence Line Analysis}
%
%
X-ray fluorescence probes physical conditions of cool circumstellar
material in the vicinity of strong X-ray sources. X-ray emission from 
a hot, optically thin photo-ionizated atmosphere occurs close to 
the illuminating X-ray source; in contrast, fluorescence from a cold,
optically thick medium is effective at orders of magnitude larger
distances. X-ray spectral properties of the Fe K region is extensively 
discussed in \citet{kallman04}. 
We detect Fe K$_{\alpha}$ and Fe K$_{\beta}$ fluorescence
lines at all phases except the high state of the main-on maximum phase
(OBSID 4585).  The fits were performed using the continuum solution
from the analysis in \S2, but with additional Gaussian line
components. We assume that line broadening is the same for both Fe
K$_{\alpha}$ and Fe K$_{\beta}$ lines. The relative wavelength
position between Fe K$_{\alpha}$ and Fe K$_{\beta} $ was fixed in the
way that the Fe K$_{\beta} $ wavelength scales by the theoretical K
fluorescence wavelengths of the neutral iron Fe I \citep{kaas93}.

The Fe K line region appears complex in almost all cases.
Besides the line fluorescence we also observe hot lines from \fexxv~
and \fexxvi~ ions (see below). Specifically the fits for the weaker Fe
K$_{\beta}$ are problematic due to its proximity to the Fe K edge
location at 1.74 \AA~ as well as the \fexxvi~line location at 1.78 \AA.  
Table \ref{tab-fe_ab} summarizes the observed values. The line
centers of the Fe K$\alpha$ line are consistent with 1.94 \AA, which
represent neutral Fe states between \fei~and \fex~. The lines appear
resolved, the widths listed in Table \ref{tab-fe_ab} are
significant. It is interesting to note that the values increase with
measured line fluxes.  Their equivalent Doppler widths range from 620
km s$^{-1}$ in the turn-off to 2100 km s$^{-1}$ in the turn-on.

The ratios of Fe K$_{\beta}$/K$_{\alpha}$
are generally larger than the theoretical value of 0.13 from \citet{kaas93}. 
More comparisons between various theoretical and experimental studies of this ratio 
are presented in \citet{palmeri03}. This study shows that such a ratio
could be as high as 0.17 for \ion{Fe}{9}. 
For our observations, during the turn-off and shortly after in the
turn-on, the values are close to the 0.13 equilibrium value. In all
other cases we observe values between 0.24 and 0.57, where the latter
seems unrealistically high.  We note that the error bars are all large
for those high ratios. Excess Fe K$_{\beta}$ might be due to errors 
in fitting the Fe K edge. In order to match the Fe K edge,
Fe K$_{\beta}$ could be over-estimated. We haven't been aware of any
intrisince atomic processes in the literature which could account
for such higher ratios.  We know that
unlike Fe K$_{\alpha}$, Fe K$_{\beta}$ must come from the less ionized
Fe ions ($<$ \fexvii) which have M shell electrons. The confusion from
the higher ionized Fe ions can only lead to the larger production of
Fe K$_{\alpha}$, which is clearly not the case here.  Therefore, the
known scenarios that possible extra contributions from the very hot
inner portion of the disk or some highly illuminated layers of the
companion star during the main-on, which all favor the larger
production of Fe K$_{\alpha}$, could be excluded. 
Thus we have no natural explanation for such a ratio

\begin{figure}[h]
 \centering
 \includegraphics[width=0.9\textwidth]{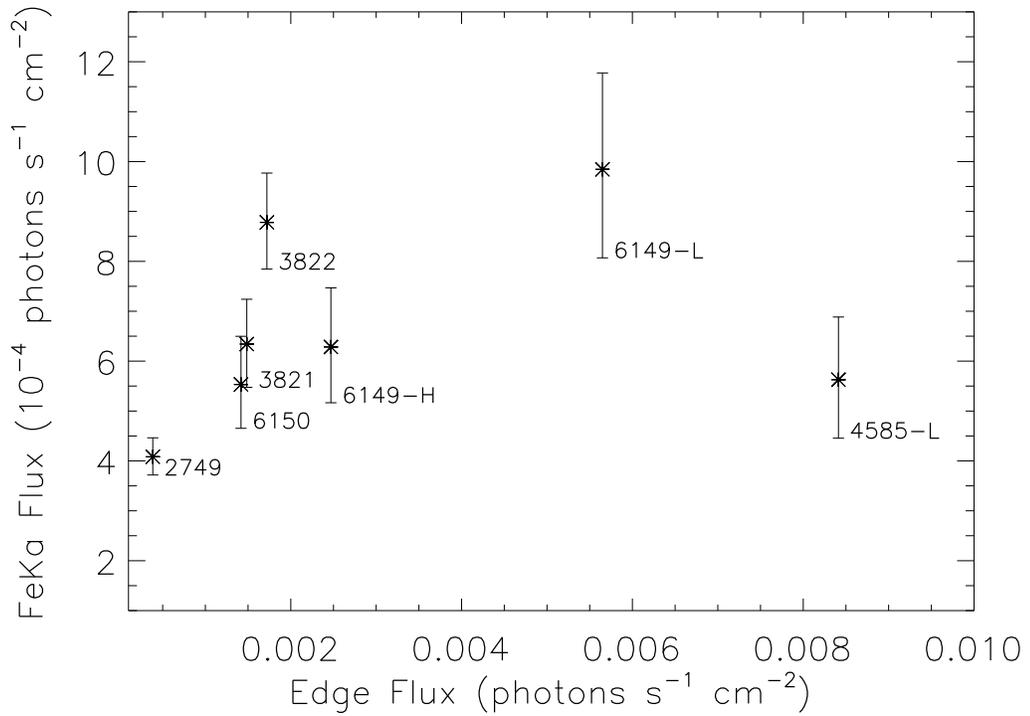}
 \caption{Fe edge flux versus Fe K$_{\alpha}$}
 \label{fig-flux-edge-feka}
\end{figure}

Figure \ref{fig-flux-edge-feka} shows the Fe K$_{\alpha}$ versus Fe K
edge flux for all observations. The edge flux is defined as the
continuum flux beyond the Fe K edge up to 1.55~\AA (8.0 keV), which
generally provides the bulk of the ionizing photons.  Similar behavior
between the Fe K$_{\alpha}$ versus Fe K edge flux has been presented
by \citet{makishima86} for a few Galactic X-ray binaries and
extragalatic AGNs. In \herx1, as
we approach the main-on, the Fe K$_{\alpha}$ line flux should increase
as the continuum flux increases. Naturally we would expect a
monotonic increase of the Fe K$_{\alpha}$ line flux with edge
flux. For most of the observations this seems to be the case.
However, this is not true for the two observations covering the
main-on low flux state.  Although here the edge flux appears twice to
three times higher with respect to the other observations, the
corresponding fluorescence line fluxes are not significantly
higher. It is to note, that fluorescence line fluxes are measured
after application of continuum column densities.  Intrinsic column
densities are much higher in the main-on low states and these 
measurements indicate less Fe K fluorescence than we would expect 
from the observed optical depth in the K edge.

\section{Line Emissivity Distributions}
We also observe highly ionized emission lines in all observations.
The existence of radiative recombination continua (RRCs) demonstrates that we observe
a photoionization dominated plasma~\citep{jimenez2005}, and their strengths are 
consistent with our measured G and  R ratios where it applies.            
No absorption features have been detected in any of the spectra.
In order to determine the ionization balance in each of the observed phases      
we also apply the photoionization code XSTAR to model all the lines.

\subsection{General Line Fits}
Using the results from the continuum analysis we first fit all
all these lines with Gaussian functions.
The discrete emission lines of H-like and
He-like ions are significantly detected at all phases except
during the high states of main-on phase (OBSIDs 4585 and 6149, for details see Table \ref{tab-lines}).
For the weak lines we fix the line width to 0.001 \AA~ and keep the line
centroid at the theoretical value. For the strong lines we fit all line components.
As it turns out, the line centroids remain with a 90\% error within
 the instrumental resolution, and line width are less than a few tens of m\AA.
No Doppler shifts have been detected.

\begin{figure}[h]
\includegraphics[width=0.5\textwidth]{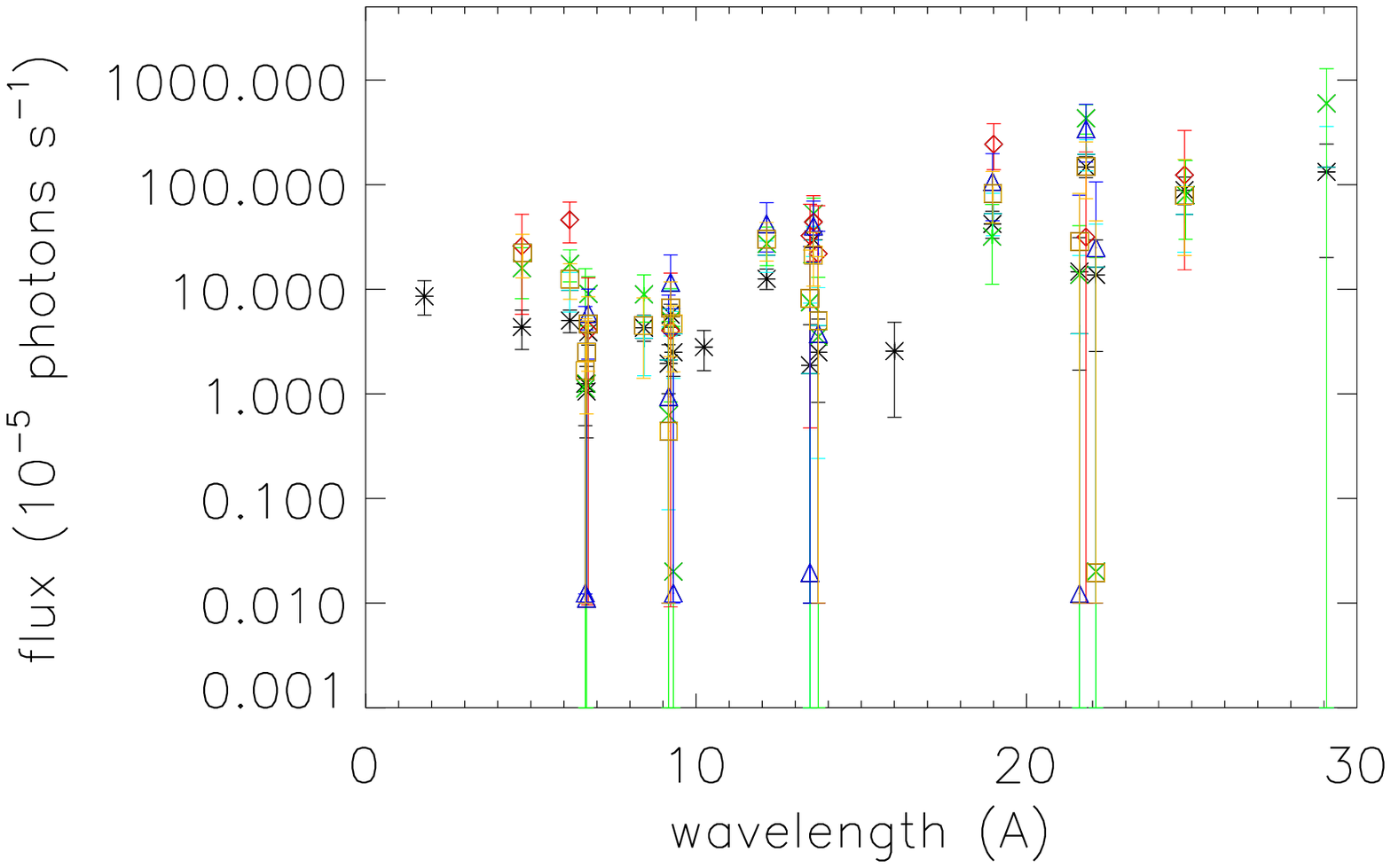}
\includegraphics[width=0.5\textwidth]{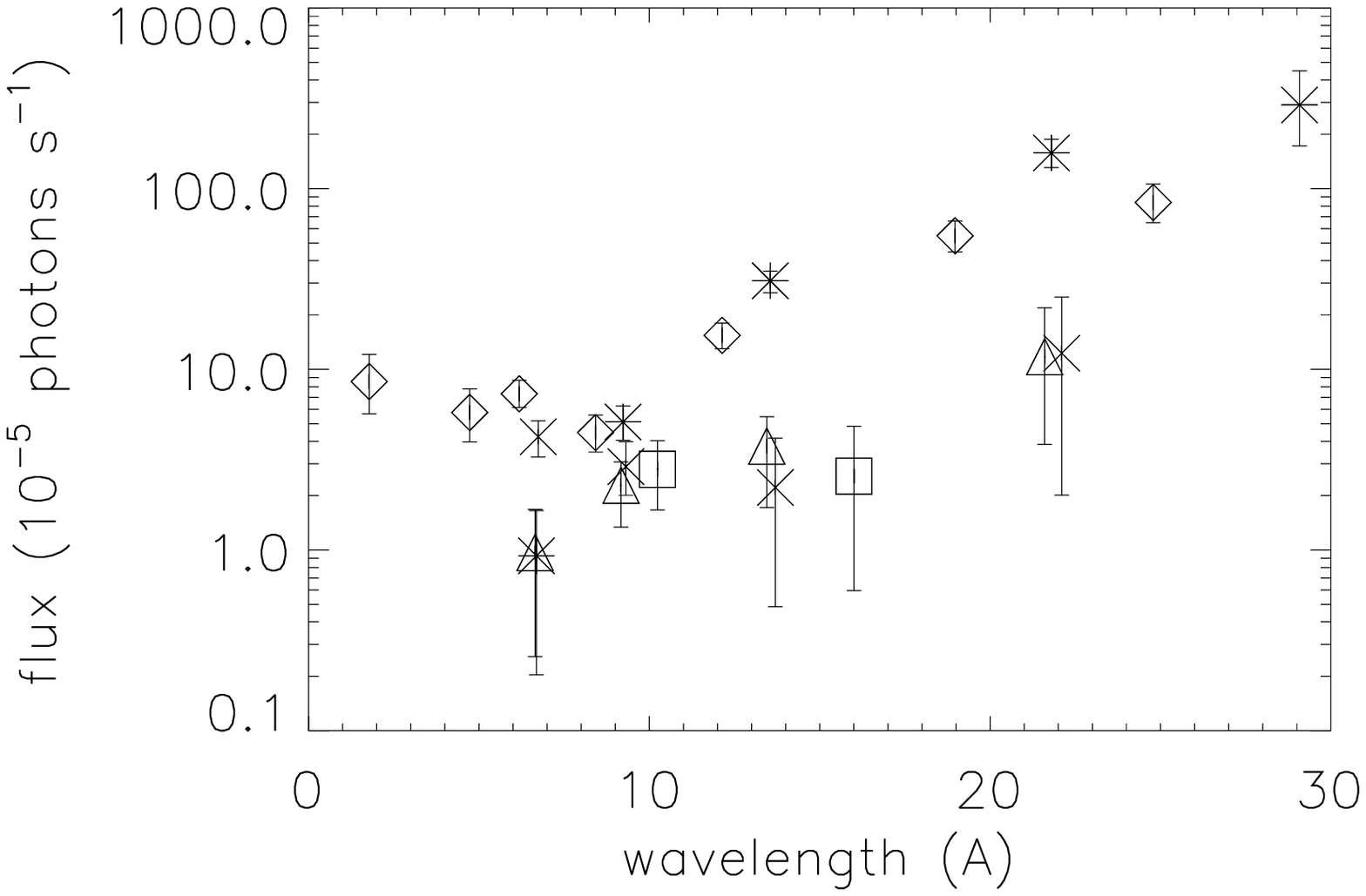}
\caption{Left: the detectable lines for all observations (ID2749-black, ID3821-cyan, ID3822-green,
               ID4585-L-red, ID6149-L-blue, ID6150-gold);
         Right: the average detectable lines (Ly$\alpha$ - diamond, Ly$\beta $ - square, 
	        resonance -triangle, intercombination - asterisk, forbidden - x ).}
\label{fig-lya}
\end{figure}

The left panel of Figure \ref{fig-lya} shows the results of the line fits for all  
observations (each color represents one individual observation). Under the assumption that
the illuminating source does not vary in between phases, we also fit the lines
simultaneously for all observations to obtain average properties. The results
of the simultaneous fits are shown in the right
panel of Figure \ref{fig-lya}. This average flux distribution gives us 
more significance for the parameters of the photoionization modeling.

\subsection{Helium-like Triplets}
%
%
Triplet lines from He-like ions are widely used to provide diagnostics \citep{porq01}
for the electron density and temperature, and also the determination
of ionization processes (photoionization or collisional ionization).
He-like lines consist of a resonance (r), an unresolved
intercombination line duplex (i), and a metastable forbidden line
(f). The most common triplet diagnostics involve ratios sensitive to
plasma density, temperature as well as ionization state. The flux ratio G = (i+f)/r is
sensitive to plasma temperature and ionization states, the flux ratio R = f/i to density.
We detected He-like triplet lines in the following states: the low
(OBSID 2749), turn-on/off states (OBSIDs 3821, 3822 and 6150), and the
main-on low states (OBSIDs 4585 and 6149), although the statistics are
poor in the most states.  No detection (less than 1 $\sigma $) is in
the main-on high states.  The fitted R and G ratios are presented in
Table \ref{tab-He}.

As claimed in \citet{baut00}, the ionization mechanism of the observed
plasma must be established before the analysis of the line
diagnostics, or these triplet line diagnostics are misleading.  RRC
are typical signatures of photoionization dominated plasma. By fitting
with the ``redge'' function in XSPEC, the electron temperature measured
from the \neix, \oviii, \ovii, and \nvii~ RRC is in the range of
several to tens of eVs, corresponding to $10^4$ to $10^6$ K.  The
broad ranged ionization states and weak RRC features thus indicate multiple temperatures
of the plasma.

In the HETG wavelength range, lines from n = 3 levels ($^{3}\rm P -
^{1}\rm S $) provide direct evidence of the photoionized nature of a
plasma \citep{baut00}. The relative strengths of these lines to the r
lines are different in coronal and photoionization plasmas.  However,
these lines have generally poor statistics in these observations and
the r line is very weak as well. 

Most of the triplet lines we detected
show strong intercombination lines, which could result in two
ways. Both strong radiation fields (by photo-excitation) and high
densities (by collisional excitation) can significantly depopulate the
$^{3}$S to $^{3}$P, leading to a strong intercombination line and a
weak forbidden line \citep{porq01}.
In case of pure photoionized plasmas where collisional excitations
occur above the critical density, we determine the density limit based
on the R ratio according to the calculations from \citet{porq00}.  R
ratios of \mgxi, \neix, and \ovii\ (see table \ref{tab-He}) are in the
range of 0.0 to 1.5, indicating densities larger than $\sim 10^{11}
~\rm cm^{-3}$, while R ratios of \sixiii ~ indicate densities of less
than $10^{14} ~\rm cm^{-3}$.

Most of the G ratios from the \sixiii, \mgxi, \neix, and \ovii~
triplets suffer from poor statistics and are given with $3\sigma$ lower
limits. The best detection is \mgxi ~ in the low state ID2749. Its
value of $4.1^{+3.9}_{-1.6} $ is consistent with the value from
\citet{jimenez2005}. The best value indicates an electron temperature
range spanning roughly from $2\times 10^{6}~\rm K $ to $5\times
10^{6}~\rm K$ \citep{porq00}.

\section{Emission Measurement Analysis}
The differential emission measure (DEM) analysis is used for 
modeling the ionization balance in X-ray binaries \citep[e.g.,][]{sako99, jimenez2005, schulz08}. 
Assuming the emission spectrum is dominated by recombination,
the line luminosity can be expressed as: 
  \begin{eqnarray}
   L_{ul} = \int n_{e}n_{z,i+1} E_{ul} \alpha^{eff} dV 
	  = \frac{A_{Z}n_{H}}{n_{e}}\int f_{i+1} E_{ul} \alpha^{eff} d (EM)
  \label{equa-EM-rr}
  \end{eqnarray}
where $A_{Z} = n_{Z}/n_{H}$ is the relative element abundance to hydrogen, 
$f_{i+1}$ is the ionic fraction of the recombining ion, $E_{ul}$ is 
the transition energy, and $\alpha^{eff}$ is the effective recombination 
rate which causes the line emission through the transition $u\rightarrow l$, 
$d(EM) =n_{e}^{2}dV$, and $n_{z,i+1}=A_{Z}n_{H}f_{i+1}$.
Given the detections of the line luminosities, emission measure (EM) 
distributions can be predicted with XSTAR. 
Note, $\alpha^{eff}$ includes the contributions from the radiative and 
di-electronic recombination, cascading effect as well as the possible
collisional excitation. As claimed in \citet[e.g.,][]{porq00}, 
the cascading contribution is significant at the low temperatures. 


We also directly derive EMs from the line formation in the XSTAR model 
without any other assumptions, which automatically accounts for the cascading effect:
  \begin{eqnarray}
  L_{ul} = \int n_{u}^{'} A_{Z}n_{H}/n_{e}^{2} E_{ul} A_{ul} d(EM)
  \label{equa-EM-cc}
  \end{eqnarray} 
 where $n_{u}^{'}= n_{u}/n_{Z}$ is the upper energy level population 
of the ion relative to the element abundance, which is easy to be extracted 
from XSTAR. $A_{ul}$ is the spontaneous transition rate. 

We first construct a reasonable grid for the photoionization plasma
applicable to \herx1. The fixed powerlaw photon index of 1.068 is
determined from the continuum fitting and the column density is low
($N_{H} \sim 10^{17}~\rm cm^{-2}$). The grid assumes three densities
($n_{e}=10^{11}, ~10^{13}, ~10^{16}~\rm cm^{-3} $), the ionization
parameter ($log~\xi $) spans from -0.1 to 5.0 with 0.1 interval
steps. All calculations assume a solar abundance.  For each of
$log~\xi$ we then derive a corresponding EM according to the
observational flux.

\begin{figure}[h]
\mbox{
\includegraphics[width=0.8\textwidth]{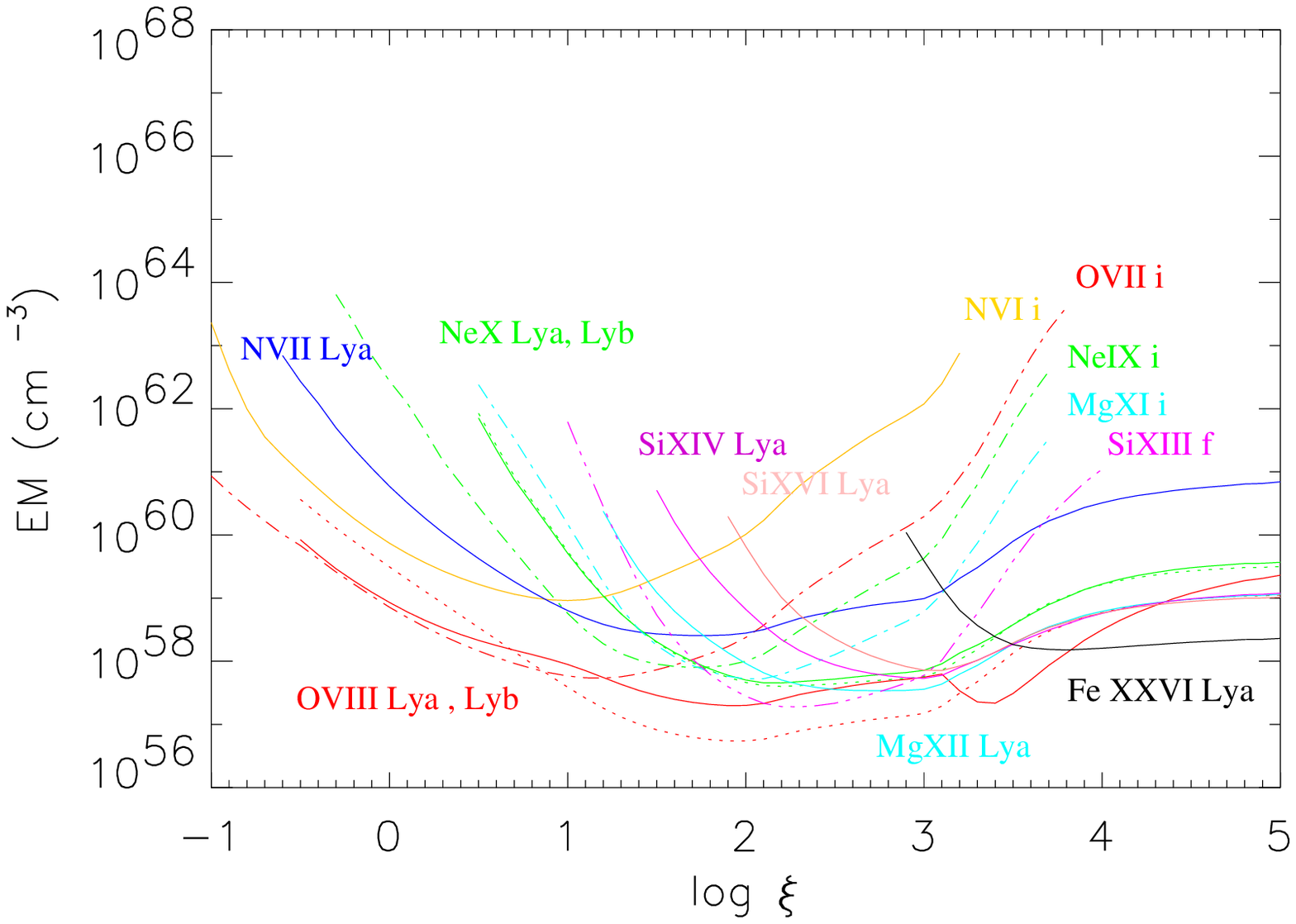}
}
\mbox{
\includegraphics[width=0.8\textwidth]{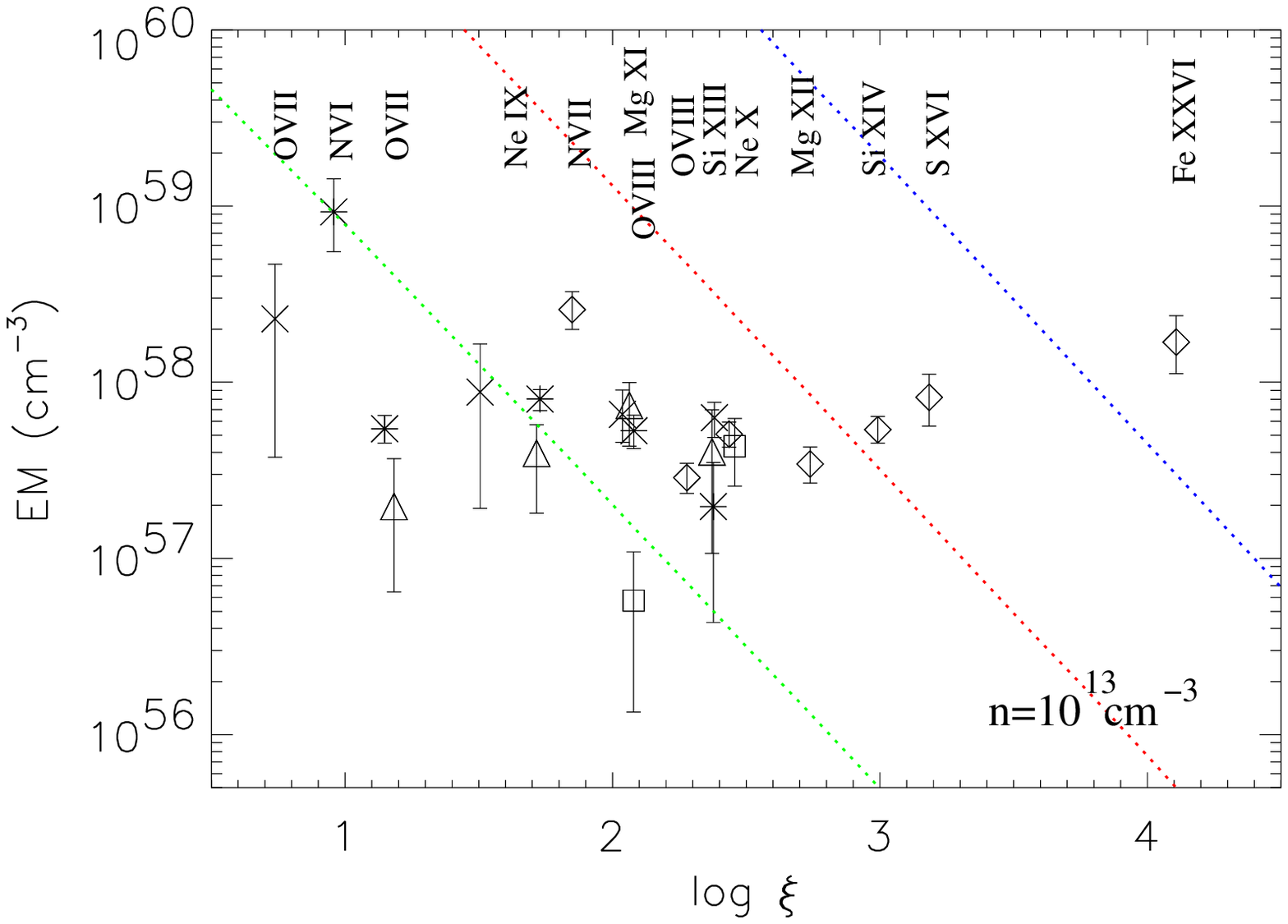}
}
\caption{{\bf{Upper}}: Emission measure curves for the strong
  lines based on the XSTAR photoionization grid.  {\bf{Lower}}:
  Symbols - the average emission measure for all detectable lines;
  symbol are the same as Figure \ref{fig-lya}.  Lines -
  calculation of emission measure in dependence of ionization
  parameter for a static corona for two different luminosities,
  $L=10^{38} \rm ~ ergs ~s^{-1}$ (blue-dash), 
  $L=10^{37} \rm ~ ergs ~s^{-1}$ (red-dash), and $L=10^{36} \rm ~ ergs
  ~s^{-1}$ (green-dash).  See the text for details.}
\label{fig-EM_obs_avg}
\end{figure}

The $n_{e}=10^{13}~\rm cm^{-3} $ photoionization XSTAR grid, for
example then determines the EM functions for the strong lines in the
upper panels of Figure \ref{fig-EM_obs_avg}.  We find that the pure
radiative recombination rate as a first approximation to the effective
recombination rate is not a good measure since the difference to the
one including cascading effect amounts to $\sim 2.5$ times 
for the OVII resonance line at $log~\xi \sim 1$.  
The cascading effect in this set of XSTAR modeling thus cannot be
neglected.

In order to get an average EM for each line, the $log~\xi$ has to be
weighted by $n_{u}A_{ul}$, which is in units of 
$\rm photons~ cm^{3}~ s^{-1}$. Specifically, we adopt the average ionization
parameter $<log~\xi> = \frac{\sum log~\xi n_{u}A_{ul}}{\sum n_{u}A_{ul}}$.
The average EM is chosen at that average ionization parameter. 
We got the average EM for all lines 
at the weighted $log~\xi$ in the lower panel of Figure \ref{fig-EM_obs_avg}.

It is noted that except for the N Ly$\alpha$ line, all the resonance
lines show a common trend in the EM distribution.  Nitrogen clearly
appears over-abundant, which is consistent with the analysis of ID2749
from \citet{jimenez2005}. They argued that 
the CNO process can produce an overabundance of nitrogen and depletion of carbon  
and oxygen for 2-12 $M\odot$ stars \citep[][and reference therein]{jimenez2005}.

The EMs of the \ovii~ and \neix ~triplets triplet lines are much more
complex. We find that the EMs of the intercombination and forbidden
lines are well above those of the resonance lines.  The reason for
this likely lies in the fact that we use one density value for a
single grid. In our case we chose a density that closely matches \mgxi
~and higher elements, which is higher than what \ovii~ and \neix~
require.  The EMs of \mgxi ~and \sixiii~ are well consistent with each
other due to the appropriate density, and the average values are about
$6\times 10^{57} ~\rm cm^{-3}$ and $4\times 10^{57} ~\rm cm^{-3}$,
respectively.  The $S^{3}$ levels in \ovii~ and \neix~ are depopulated
in excess, which results in higher emissivities (therefore the higher EM)
of the intercombination lines.  The calculations show that we get
fairly matching EMs for \ovii~ at a density of $\sim 1\times 10^{12}
~\rm cm^{-3}$.  The forbidden line EM never fully match which is
likely a consequence that we seem to observe higher fluxes than
predicted from the emissivities.

Another effect we observe for the \ovii~ and \neix~ triplets is that
their weighted $log ~\xi$ are slightly different.  The weighted
$log~\xi$ of the forbidden lines are much lower than that of the
intercombination and resonance lines, particularly for \ovii.  We
think this is due to the different atomic transition processes
accounted for in XSTAR with respect to these triplet lines.  There are
three atomic processes which can excite the He-like ion in a
photoionized plasma: 1) direct collisional excitation from the ground
state as seen in the collisional plasmas; 2), either radiative or
dielectronic recombination from H-like ions; and 3), K shell
photoionization of Li-like ions.  Process (1) excites the various
upper levels, mainly 1s2p(1P), resulting in the resonance line.  Since
it is generated from the ground state, it traces the He-like ion
abundance.  Process (2) excites the 1s2p(3P) more than the 1s2s(3S)
and 1s2p(1P). The radiative recombination tends to populate 
the upper levels according to 3:1:1, which however is modified by
the 3P branching to 2.1:1.9:1.0 . 
This process involves the H-like ion, so it shifts the
emissivity curves for the intercombination line to the higher
$\xi$. Therefore the weighted $\xi$ of the intercombination line is
higher than that of the resonance line.  Process (3) only excites the
1s2s(3S). It comes from a Li-like ion, tracing the abundance of the
\ovi~ ion, and shifting the forbidden lines to the lower
$\xi$. Therefore, the weighted $\xi$ of the forbidden line is lower than
that of the resonance line.

We also note that in the current XSTAR version, K shell
photoionization of the Li-like ion for \neix\ has not been
updated and may not be as accurate as the other ions. 
For O and Fe, the XSTAR K shell  photoionization cross sections 
from Li-like ions are taken from R-matrix calculations \citep{Garcia05,baut03}, 
and so are likely to be an accurate description of the distribution of final
states in the He-like ions.  For other elements, the K shell cross sections 
are taken from central potential calculations \citep{verner95}
and assume that all the final state ions are produced in their ground state.  
Thus they do not account for the production of He-like ions in the 1s2s(3S) state.
Therefore, the atomic process (3) cannot fully account for the lower $\xi$ for
\neix ~at its appropriate density of $3\times10^{12}\rm ~cm^{-3}$.  We
speculate that it is a temperature effect that causes the different
log $\xi$'s for \neix. The resonance and intercombination lines
usually favor higher temperatures leading to higher ionization
parameter estimates.  Last, but not least, we should emphasize that
assigning a single $\xi$ to the \ovii~ and \neix ~ triplets is not
ideal. Future approaches should feature deconvolution techniques to
trace different ion abundances once better physical models become
available for Her~X-1.




As argued by \citet{schulz08} (Figure 5), for \herx1 with a neutron
star about $1.5\pm 0.3 ~\rm M_{\odot}$, $\frac{dEM}{d\xi}\sim
2.0\times 10^{65}\xi^{-3}L^{2}_{38}R_{i7}^{-3/2}T_{4}^{1/2}$ for a
static corona. However, here we use the more accurate numerical
integration of eq.(2) in \citet{schulz08}, and these are shown
as the dash lines in Figure \ref{fig-EM_obs_avg}
for a single-stream treatment of the transfer of the X-ray
luminosity of $L=10^{38}~\rm (blue-dash), 10^{37}~\rm (red-dash), 
and~ 10^{36}~\rm(green-dash) ~ ergs ~ s^{-1}$ 
for a disk extending from $10^{8.3}$ to $10^{11}\rm ~cm$. Note, we set
the inner disk radius to the corotation radius: $r_{co}=
(GMP^{2}_{spin}/4\pi^{2})^{1/3} \sim 2\times 10^{8} \rm ~ cm$, which
is also about the magnetosphereic radius. This shows that 
at $log~\xi\sim4.1$, an emission measure ($1.7\times 10^{58} \rm
cm^{-3}$ for \fexxvi, for example) that we infer from the observations
can only be produced by an ADC if the source luminosity is well above
$10^{38} \rm ~ ergs ~s^{-1}$.  However, the luminosity we infer from
the continuum fits is almost one order of magnitude lower. 
Several other lines fluxes (\sixiii ~ and \sxvi) exceed a source 
luminosity of $10^{37} \rm ~ ergs ~s^{-1}$ although they are only 
slightly higher than the predictions of the static corona. 
Similar for Cir X-1, the true luminosity intrinsic to the source may 
be much greater than what we observe, which might be due to the 
obscuration or collimation of the radiation into a direction away 
from our line of sight.

\section{Discussion}

Energy spectra early on indicated a dominant presence of photoelectric
absorption and electron scattering during most states in
\herx1~\citep{becker77, davison77, pravdo77, parmar80}. Column
densities deduced from broadband low resolution spectrometers range
from $< 7\times 10^{21}~\rm cm^{-2}$ in the on-state to $3\times
10^{23}~\rm cm^{-2}$ in dips. Thus, the Galactic column of about 
$10^{18}~\rm cm^{-2}$ is negligible in
comparison to those intrinsic values between $10^{21}$ and $10^{23}~\rm
cm^{-2}$. With the availability of increased spectral resolution and
sensitivity, more detailed information about the nature of the
spectral continuum emerged at specific orbital and superorbital
phases.  Rocket flights in the mid-1970s discovered significant soft
emission coinciding with high flux periods and low columns
\citep{shulman75, catura75} which has later been confirmed as soft
($\sim 0.1$ keV) blackbody emission by \einstein~\citep{mccray82} and
\sax~\citep{oosterbroek97}.  \xmm observed \herx1 at three 
superorbital phases, near main-on, short-on, and low state, and
modeled the continuum with a partial covering absorption model which
included the powerlaw and as well as the soft blackbody
spectrum. Extreme changes in partial column were observed, ranging from
the low interstellar value near the main-on to $6\times 10^{23}~\rm
cm^{-2}$ for the powerlaw component in the low state, with significant
changes of the ratio of the covering fractions between
states~\citep{ramsay02}. These latter authors also find that by keeping the powerlaw
index fixed to 0.85 the fits above 2 keV also matched consistently
well.  While closely monitoring the turn-on phase with \rxte
~and expanding the energy range to 17 keV, \citet{kuster05} could further
develop the partial covering model above 2 keV involving a fixed
powerlaw index of 1.068 and an evolution of cold photoelectric
absorption and Thomson electron scattering.

This partial covering model involving a soft blackbody and a cut-off
powerlaw with fixed index at 1.068 and a cut-off at 21.5 keV works
well for all HETG observations between about 0.3 and 8 keV. The
blackbody, with an average temperature of 114$\pm32$ eV, is required
at all phases including the highly absorbed low states.  The emission
radius amounts to somewhat less than 100 km, which is smaller than the
one deduced from the \xmm observations~\citep{ramsay02}, but still
large enough to favor an association with the inner disk over the
neutron star surface.  Column densities appeared consistently high and
highly variable --- varying from $9.3\times 10^{22}~\rm cm^{-2}$
during the the Main-high state to $3.2\times 10^{23}~\rm cm^{-2}$
during the Main-low state.  The fact that we observe sizeable
absorption even during the high flux states is remarkable,
specifically since \xmm observations do not find significant amounts
during the high flux states that they viewed \citep{ramsay02}.
However, our observed absorption is consistent with the findings by
~\citet{kuster05} during their monitoring of the turn-on. The Chandra
HETG archive observations cover  several different orbital and 
superorbital phases of \herx1. The fact that we obtain good fits with a
nearly fixed continuum shape (frozen powerlaw slope and reasonably constant
blackbody temperature) and minimal observable changes in the line derived 
ionization balance, while we see large changes in the column, leads us to the conclusion that
most if not all changes in X-ray flux are due to absorption,
obscuration, and scattering effects relating to line of sight geometry.

Strong line fluorescence is a good indicator of large amounts of
cool circumstellar material present in close vicinity of the central
X-ray source. Iron line fluorescence was discovered in \herx1
by~\citet{pravdo77} and not only is it present
at all super-orbital phases~\citep{kahabka95} but there are indications
that it varies over the
1.24 second pulse phase and the 35 day cycle~\citep{choi94}. We also
resolve Fe line fluorescence in most observations, but only once during the
Main state. This shows that fluorescence is not detectable at all
time. Furthermore, it was not detected during Main-high where the
X-ray flux was highest. This is actually consistent with another
finding of our study. Figure 6 shows a positive correlation of the Fe
K$_{\alpha}$ line flux with the source flux, but some anti-correlation
with equivalent width. The non-detection of the line during Main-high
is equivalent to a non-detection of any equivalent width, which fits
the same trend. The anti-correlation resembles the ``Baldwin effect''
which was originally based on a decrease of \ion{C}{4} EWs with respect to UV
luminosity in AGN~\citep{baldwin77}.  Its physical basis is likely a
progressing ionization of cold to warm material columns with
increasing luminosity~\citep{nayakshin00}. Similar trends have been
observed in \exo and \chandra data of X-ray
binaries~\citep{gottwald91, torrejon09}.  The \xmm observations seem
to follow this trend as well, but with a major difference that there a
broad line was observed with a line centroid shifted towards higher
ion states~\citep{ramsay02}.  This is an important result because it
shows that if we associate the line fluorescence with cool circum-disk
material, the outer disk is either not fully responsible for all the
absorption and obscuration observed or the X-ray source itself is
changing. In conjunction with the above conclusion that the unabsorbed
spectral shape remains more or less constant throughout all the flux
changes, any intrinsic changes to the source are solely due to changes
in the normalization of the spectrum.
We observe Fe K$_{\alpha}$ line broadenings with
velocities between 600 and 2000 km s$^{-1}$ consistent with the \xmm
measurements, however, we clearly do not observe any significant line shifts.
Given the lack of energy shifts, the line broadening is likely of
dynamic origin and not associated with either gravitational effects or emission
from highly ionized Fe states, as might be expected from the very  
innermost regions of the disk. A major contribution of the optically thick part 
of the disk to the
line fluorescence seems unlikely because the line properties seem
not well correlated with orbital and superorbital phases.
Detailed geometrical and optical modeling is necessary, 
however, to arrive at a consistent picture for the source of this fluorescent emission.
At this point we favor the view that dense cool material is
scattered within the corona, i.e., over radii $\approxgt 10^{9} ~\rm cm$, 
consistent with the measured line widths.

The possible increasing presence of warm material at higher observed source fluxes 
may have another consequence. We observe ratios of the Fe K$_{\beta}$
over Fe K$_{\alpha}$ line fluxes which are only near the equilibrium
value of 0.13 in the low state and in the turn-on from a low state. In
all the other states the ratio is substantially larger, although the
corresponding error bar is larger too.  One effect we have
to consider is that the Fe K$_{\beta}$ is relatively weak and blended
on top of the Fe K edge. At the higher continua contrast is decreased
and the line flux may be overestimated.  Although there is no direct
indication of the existence of warm columns other than through the
Baldwin effect, part of the high line ratios could be due to blends with
unresolved warm transitions.

The main goal of this study is to investigate the line emissivity
distributions at various orbital and superorbital phases in
\herx1. 
In three \xmm RGS observations \citet{jime02}
covered the low, short-on and main state. In those observations the
low and short-on flux states produced emission lines from carbon to
neon and produced CNO abundances and ADC properties. They also
reported an absence of ADC line emission during the main state. An
extended ADC was identified in the \herx1 low state by
\citet{jimenez2005} using \chandra HETG spectra. Our study is an
expansion of this study. We detect ADC line emission in the low, the
turn-on, and in the main-dip and main-low states.  We do not detect
lines emission during the main-high state consistent with the findings
by \citet{jime02}.  Figure 7 plots all the unabsorbed line fluxes we
detected at these states. The flux distribution indicates that the
measured ADC lines are likely part of a common EM distribution. Line
fluxes vary within a factor two or three between states, in which
continuum fluxes changes by over an order of magnitude. This leads us
to conclude that most changes in the line flux are due to some
residual obscuration rather than physical changes of the line emitting
region. This is consistent with the above conclusions. It also
means that neither the central source nor the ADC emissivities are
changing with orbital and superorbital phase.

We therefore calculate an average emissivity distribution from all the
observations. This resulting distribution is very similar to the one
deduced by ~\citet{jimenez2005}. We also find that nitrogen is
generally overabundant, which Jimenez-Garate et al. explain in
conjunction with the observed carbon depletion in the RGS observations
\citep{jime02} and HZ Her's CNO cycle. From the R ratios as well as
the EM distribution we infer an ADC density of $1\times10^{13} \rm ~
cm^{-3}$ for \ion{Mg}{11}. From emission measure adjustments of
\ion{Ne}{9} and \ion{O}{7} we also deduce densities of $3\times10^{12}
\rm  ~cm^{-3}$ and $1\times10^{12} \rm  ~cm^{-3}$, respectively.  Lower
densities at lower Z elements are not unusual because their ionization
regimes likely involve wider ranges of radii and ionization
parameters. The size of the accretion disk is thought to be a few
times $10^{11}$ cm~\citep{cheng1995}, which puts a lower limit to
Doppler broadening to be observed in the lines. There is likely
a contribution from UV flux locally to the disk which also affects
the R value. Significant UV flux would imply that our inferred 
densities are the upper limits. 
In case of the Si XIII triplet we observe a large R ratio which
neither the UV nor the collisional transitions can explain. For this
ion, the implied density is then a valid upper limit.
Our observed line
velocities range between 200 -- 800 km s$^{-1}$, which assuming an
orbital inclination of 81 deg~\citep{howarth83} results in ADC radii
between $2\times10^{10} ~\rm cm \approxlt r \approxlt
3\times10^{11}~\rm cm$. With respect to the findings by
\citet{jimenez2005} this is very similar except for a smaller inner
radius. This is because we find significantly higher velocities at
higher flux states. These radii are consistent with ADCs found sources
like 4U 1822-37~\citep{cottam01}, 2S 0921-63~\citep{kallman03}, and
Cir X-1~\citep{schulz08}. These radii are significantly larger than
the ones found recently in the Z-source Cyg X-2~\citet{schulz09} which
features much higher source luminosities and Doppler line velocities.

We also tested how consistent our observed line emissivities are with
respect to the observed source luminosity and a static ADC. Using the
model calculations presented by ~\citet{schulz08}, we computed
emission measures as a function of ionization parameters for various
source luminosities. While most low energy lines can be well-explained
by the observed source luminosity of $5\times10^{37}$ \ergsec, the 
observed H-like lines of \ion{S}{16} and \ion{Fe}{26} are not explained 
by such a flux. Such deficiencies,
i.e., significant highly ionized lines for a moderate flux, have been
found in ADCs of 4U 1822-37~\citep{cottam01}, Cir
X-1~\citep{schulz08}, and here for \herx1.  Further ADC modeling, with
a detailed focus on heating budgets, is necessary to find an
explanation for these H-like lines. The discrepancies from simple
models, however, may be too high to be addressed by simply modifying
the static corona model.  More likely candidates may need to invoke
additional sources of energy, such as magnetic field activities, to
explain these results.
       
\begin{center}
ACKNOWLEDGMENTS
\end{center}

We thank R\"{u}diger Staubert for his help with the ephemeris, and Mike Noble for his help 
with running the XSTAR models. We gratefully acknowledge the financial support 
of Chandra X-Ray Observatory theory grant TM8-9005X.
%
%

%
%
\begin{deluxetable}{l|ccccc}
\tiny
\tablecolumns{6}
\tablecaption{Observations for \herx1}
\tablewidth{0pt}
\tablehead{
obsID & MJD interval  & Obs. Start & Exp.(ks) & $\theta_{Orb.}$\tablenotemark{a} & $\Psi_{35d} $\tablenotemark{a} }
\startdata
2749 &    52399.427 - 52400.032 & 2002-05-05 10:14:49 UT & 50.17& 0.33 - 0.68 &  0.42 \\
3821 &    52875.886 - 52876.261 & 2003-08-24 21:15:16 UT & 30.10& 0.57 - 0.79 &  -0.03\\
3822 &    52982.987 - 52983.379 & 2003-12-09 23:41:44 UT & 30.14& 0.57 - 0.80 &  0.04 \\
4585 &    53335.256 - 53335.512 & 2004-11-26 06:08:16 UT & 20.16& 0.76 - 0.91 &  0.10 \\
6149 &    53338.630 - 53338.913 & 2004-11-29 15:06:38 UT & 22.14& 0.75 - 0.91 &  0.19 \\
6150 &    53340.370 - 53340.647 & 2004-12-01 08:52:39 UT & 22.05& 0.77 - 0.93 &  0.24 \\
\enddata
\label{tab-obsid}
\tablenotetext{a}{Using $P_{orb}=1.700167387 \rm~d$ at $T_{\pi/2}=50290.659202 \rm~(MJD)$ (Private contact from R. Staubert)}
\end{deluxetable}

\begin{landscape}
%
%
\begin{center}
\begin{deluxetable}{l|cc|cc|c|c|c|c}
\tiny
\tablecolumns{9}
\tablecaption{Continuum fits for \herx1 using pcfabs(bbody+cutoffpl)}
\tablewidth{0pt}
\tablehead{
obsID & nH               & $\rm f_{pc}$ &    kT & K1\tablenotemark{a}                       & K2\tablenotemark{b} &  $\chi^{2}$/dof & Flux\tablenotemark{d} &Lumi\tablenotemark{e} \\
      & $10^{22}cm^{-2} $& & (keV) & $10^{39} \rm~ ergs~ s^{-1}/(10~ kpc)^{2}$ & $\rm photons ~keV^{-1}~ cm^{-2} ~s^{-1}$ & &     
}
\startdata
2749    & $19.6^{+1.6}_{-1.5} $\tablenotemark{c} & $0.753^{+0.020}_{-0.018}$ & $0.129^{+0.012}_{-0.012}$  & $1.3^{+0.4}_{-0.3}\times 10^{-4}$  & $7.4^{+0.7}_{-0.6}\times 10^{-3}$  &1026/598 & 0.4 &0.5   \\
\cline{1-9}
3821    & $19.3^{+1.4}_{-1.3} $                  & $0.833^{+0.008}_{-0.009}$ & $0.100^{+0.014}_{-0.013}$  & $4.8^{+2.7}_{-1.6}\times 10^{-4}$  & $3.1^{+0.2}_{-0.2}\times 10^{-2}$  &1068/781 & 1.0 &1.9  \\
3822    & $28.3^{+1.7}_{-1.4} $                  & $0.907^{+0.005}_{-0.005}$ & $ > 0.052$  & $> 3.3\times 10^{-4}$  & $4.4^{+0.3}_{-0.2}\times 10^{-2}$  &970/673 & 0.7 &2.8  \\
\cline{1-9}
4585-H  & $11.3^{+1.2}_{-1.1} $                  & $0.734^{+0.012}_{-0.012}$ & $0.141^{+0.008}_{-0.008}$  & $3.5^{+0.5}_{-0.4}\times 10^{-3}$  & $1.4^{+0.1}_{-0.1}\times 10^{-1}$  &977/795  & 9.9 &9.7\\
6149-H  & $10.3^{+1.0}_{-0.6} $                  & $0.699^{+0.004}_{-0.007}$ & $0.136^{+0.003}_{-0.004}$  & $3.1^{+0.3}_{-0.2}\times 10^{-3}$  & $9.4^{+0.3}_{-0.2}\times 10^{-2}$  &1665/1208& 8.8&6.6\\
\cline{1-9}
4585-L  & $30.0^{+0.8}_{-0.8} $                  & $0.95                   $ & $0.110^{+0.015}_{-0.016}$  & $1.7^{+1.1}_{-0.6}\times 10^{-3}$  & $7.9^{+0.2}_{-0.2}\times 10^{-2}$  &664/454& 0.9 &5.2 \\
6149-L  & $24.6^{+1.5}_{-1.4} $\tablenotemark{c} & $0.884^{+0.010}_{-0.010}$ & $0.127^{+0.014}_{-0.013}$  & $1.2^{+0.4}_{-0.3}\times 10^{-3}$  & $5.8^{+0.5}_{-0.5}\times 10^{-2}$  &520/426& 1.6& 3.8\\
6150    & $27.7^{+2.3}_{-1.9} $                  & $0.897^{+0.008}_{-0.007}$ & $0.110^{+0.014}_{-0.013}$  & $6.4^{+3.0}_{-2.0}\times 10^{-4}$  & $3.6^{+0.3}_{-0.2}\times 10^{-2}$  &782/530  & 0.8 &2.3\\
\enddata
\tablenotetext{a}{Normalization for blackbody component}
\tablenotetext{b}{Normalization for cutoff-powerlaw component, where photon index is fixed at 1.068 and high energy cut at 21.5 keV}
\tablenotetext{c}{at 67\% confidence level}
\tablenotetext{d}{detected flux is in unit of $10^{-11}\rm ~ergs ~ s^{-1}~cm^{-2}$ and in [0.3,10] keV band}
\tablenotetext{e}{absorption corrected luminosity is in unit of $10^{36}\rm ~ergs ~ s^{-1}$ and in [0.3,10] keV band}
\label{tab-con}
\end{deluxetable}
\end{center}
%
%
\begin{center}
\begin{deluxetable}{l|l|cc|cc|c|c}
\scriptsize
\tablecolumns{8}
\tablecaption{Continuum fits for \herx1 using pcfabs(bbody+cutoffpl)}
\tablewidth{0pt}
\tablehead{
Stage&obsID & nH               &$\rm f_{pc}$&    kT & K1\tablenotemark{a}                       & K2\tablenotemark{b} &  $\chi^{2}$/dof \\
  &    & $10^{22}cm^{-2} $& & (keV) & $10^{39} \rm~ ergs~ s^{-1}/(10~ kpc)^{2}$ & $\rm photons ~keV^{-1}~ cm^{-2} ~s^{-1}$ & 
}
\startdata
Low &2749    & $19.6^{+1.6}_{-1.5} $\tablenotemark{c} & $0.753^{+0.020}_{-0.018}$ & $0.129^{+0.012}_{-0.012}$  & $1.3^{+0.4}_{-0.3}\times 10^{-4}$  & $7.4^{+0.7}_{-0.6}\times 10^{-3}$  &1026/598\\
\cline{1-8}
Turn-on &3821    & $19.3^{+1.4}_{-1.3} $                  & $0.833^{+0.008}_{-0.009}$ & $0.100^{+0.014}_{-0.013}$  & $4.8^{+2.7}_{-1.6}\times 10^{-4}$  & $3.1^{+0.2}_{-0.2}\times 10^{-2}$  &1068/781  \\
Turn-on &3822    & $28.3^{+1.7}_{-1.4} $                  & $0.907^{+0.005}_{-0.005}$ & $ > 0.052$  & $> 3.3\times 10^{-4}$  & $4.4^{+0.3}_{-0.2}\times 10^{-2}$  &970/673  \\
\cline{1-8}
Main-high  &4585,6149& $9.3^{+1.6}_{-0.8} $                  & $0.631^{+0.011}_{-0.007}$ & $0.143^{+0.005}_{-0.004}$  & $37.7^{+3.9}_{-2.5}\times 10^{-4}$  & $11.3^{+0.6}_{-0.3}\times 10^{-2}$  &1352/1052\\
Main-dip  &4585,6149& $9.8^{+0.7}_{-0.7} $                  & $0.769^{+0.008}_{-0.008}$ & $0.124^{+0.007}_{-0.007}$  & $7.9^{+2.2}_{-2.2}\times 10^{-4}$  & $12.8^{+0.5}_{-0.5}\times 10^{-2}$  &1060/843\\
Main-low  &4585,6149& $32.2^{+1.2}_{-1.2} $                  & $0.946^{+0.002}_{-0.003}$ & $0.121^{+0.012}_{-0.011}$  & $1.1^{+0.3}_{-0.3}\times 10^{-3}$  & $8.2^{+0.4}_{-0.4}\times 10^{-2}$  &1051/670\\
Turn-off &6150    & $27.7^{+2.3}_{-1.9} $                  & $0.897^{+0.008}_{-0.007}$ & $0.110^{+0.014}_{-0.013}$  & $6.4^{+3.0}_{-2.0}\times 10^{-4}$  & $3.6^{+0.3}_{-0.2}\times 10^{-2}$  &782/530  \\
\cline{1-8}
Main-Ht  &4585,6149& $15.3^{+2.0}_{-1.8} $                  & $0.640^{+0.007}_{-0.010}$ & $0.143^{+0.004}_{-0.004}$  & $38.6^{+2.6}_{-2.7}\times 10^{-4}$  & $11.6^{+0.5}_{-0.5}\times 10^{-2}$  &1350/1052\\
Main-Dt  & 4585,6149& $15.2^{+1.2}_{-1.1} $                  & $0.769^{+0.008}_{-0.008}$ & $0.124^{+0.006}_{-0.005}$  & $7.9^{+2.2}_{-2.2}\times 10^{-4}$  & $12.8^{+0.5}_{-0.4}\times 10^{-2}$  &1059/843\\
Main-Lt  & 4585,6149& $50.2^{+1.9}_{-1.9} $                  & $0.947^{+0.002}_{-0.002}$ & $0.121^{+0.012}_{-0.011}$  & $1.1^{+0.3}_{-0.3}\times 10^{-3}$  & $8.2^{+0.4}_{-0.4}\times 10^{-2}$  &1045/670\\
\enddata
\tablenotetext{a}{Normalization for blackbody component}
\tablenotetext{b}{Normalization for cutoff-powerlaw component, where photon index is fixed at 1.068 and high energy cut at 21.5 keV}
\tablenotetext{c}{at 67\% confidence level}
\label{tab-con1}
\end{deluxetable}
\end{center}
\end{landscape}

\begin{deluxetable}{l|ccccccc}
\tiny
\tablecolumns{8}
\tablecaption{FeK$_\alpha$, FeK$_\beta$ for \herx1 with pcfabs(bbody+cutoffpl)}
\tablewidth{0pt}
\tablehead{
obsID  & $\lambda_{\rm K_{\alpha}}$   & F($\rm K_{\alpha}$)\tablenotemark{a}  & EW($\rm K_{\alpha}$) &$\sigma_{\rm K_{\alpha},_{\beta}}$   &  R\tablenotemark{b}=$\rm \frac{F(K_{\beta})}{\rm F(K_{\alpha})}$  &  Cash/dof & $\tau $\\
       & ($\AA$)                   & ($10^{-4}$) & (eV) &($10^{-3}\AA$ )      &     &  &0.1 }
\startdata
2749 & $1.938^{+0.001}_{-0.001} $ & $4.1^{+0.4}_{-0.4}$  &$635^{+59}_{-57}$ & $2.8^{+1.2}_{-1.6} $ & $0.16^{+0.07}_{-0.06}  $ &  52.3/26 & $<1.7$\\
\cline{1-8}
3821 & $1.939^{+0.001}_{-0.002} $ & $6.3^{+0.9}_{-0.9} $ &$243^{+34}_{-33}$ & $9.9^{+2.7}_{-2.3} $ & $0.19^{+0.13}_{-0.12} $  &  62.6/28 & $3.0^{+1.6}_{-1.5}$\\
3822 & $1.938^{+0.001}_{-0.001} $ & $8.8^{+1.0}_{-0.9} $ & $274^{+31}_{-29}$ &$7.5^{+1.6}_{-1.4} $ & $0.38^{+0.12}_{-0.11} $  &  94.0/28 & $5.7^{+1.6}_{-1.5}$\\
\cline{1-8}
4585-H & --- & --- & --- & --- & --- &  --- &---\\
6149-H & $1.940^{+0.001}_{-0.001}$ & $6.3^{+1.2}_{-1.1} $ & $81^{+15}_{-14}$   &1.0 & $0.24^{+0.22}_{-0.18} $  & 236.7/28 &---\\
\cline{1-8}
4585-L & 1.94 & $5.7^{+1.3}_{-1.2} $ &$116^{+26}_{-24}$ & 1.0 & $0.57^{+0.30}_{-0.23} $  &  226.0/29 & $5.5^{+1.5}_{-1.4}$\\
6149-L & $1.940^{+0.001}_{-0.001}$ & $9.8^{+1.9}_{-1.8} $ & $207^{+41}_{-37}$ & $7.5^{+2.7}_{-2.1}$ & $0.23^{+0.21}_{-0.17} $  & 55.9/26 & $5.7^{+2.6}_{-2.4} $\\
6150 & $1.939^{+0.001}_{-0.002} $ & $5.5^{+1.0}_{-0.9} $ & $209^{+37}_{-33}$ & $5.1^{+3.0}_{-2.9} $ & $0.35^{+0.18}_{-0.15} $  &  76.4/28 & $3.3^{+1.7}_{-1.6}$
\enddata
\label{tab-fe_ab}
\tablenotetext{a}{Its unit is $\rm photons ~s^{-1} ~cm^{-2} $}
\end{deluxetable}

\begin{landscape}
\begin{deluxetable}{l|c|c|cc|cc|ccc}
\tiny
\tablecolumns{10}
\tablecaption{line flux\tablenotemark{a} detections for \herx1}
\tablewidth{0pt}
\tablehead{
lines & $\lambda$& F(low state) &\multicolumn{2}{c}{F(main on)}&\multicolumn{2}{c}{F(main high)}&\multicolumn{3}{c}{F(main low)} \\
      & ($\AA$) & 2749 & 3821&3822& 4585H & 6149H &4585L &6149L&6150}
\startdata
 $\rm S~XVI~Ly\alpha$ &   4.729  & $4.4^{+2.0}_{-1.7}$  &--- &$15.9^{+9.1}_{-7.8}$ &--- & ---&$25.9^{+26.1}_{-20.1} $& --- &$22.3^{+1.1}_{-9.4}$ \\ 
 \cline{1-10}
 $\rm Si~XIV~Ly\alpha$&   6.185  & $5.0^{+1.3}_{-1.2}$  & $9.8^{+4.6}_{-3.8}$& $17.5^{+6.4}_{-5.7}$ & ---& &$46.1^{+22.0}_{-18.4}$& ---&$12.5^{+5.1}_{-4.5}$\\
 $\rm Si~XIII~r$&         6.648  & $1.3^{+0.8}_{-0.8}$  &--- &$<14.6$&---&---&---&$<6.9 $&$<3.8$\\
 $\rm Si~XIII~i$&         6.686  & $1.1^{+0.8}_{-0.7}$  &--- &$<3.2$ &---&---&---&$<4.1 $&$2.5^{+5.0}_{-0.6}\tablenotemark{b} $\\
 $\rm Si~XIII~f$&         6.741  & $3.9^{+1.1}_{-1.0}$  &--- &$9.0^{+4.2}_{-3.5}$ &---&---&$<12.9 $&$5.9^{+4.2}_{-3.7}\tablenotemark{b}$&$4.7^{+3.8}_{-3.0}$\\
 \cline{1-10}
 $\rm Mg~XII~Ly\alpha$&   8.423  & $4.3^{+1.3}_{-1.1}$  & $3.4^{+2.2}_{-1.9}$& $8.9^{+4.8}_{-4.1}$ &---&---&---&---&$4.5^{+3.7}_{-3.1}$ \\ 
 $\rm Mg~XI~r$&           9.169  & $1.9^{+1.0}_{-0.9}$  & $2.1^{+2.3}_{-2.0}$&$ <4.3$ &---&---&---&$<8.8 $&$<3.0$\\
 $\rm Mg~XI~i$&           9.229  & $5.7^{+1.5}_{-1.4}$  & $6.5^{+3.3}_{-2.9}$&$5.0^{+5.1}_{-4.2}$ &---&---&$<14.2 $&$11.6^{+9.7}_{-7.8} $&$6.6^{+5.1}_{-4.4}$\\
 $\rm Mg~XI~f$&           9.314  & $2.5^{+1.1}_{-1.0}$  & $3.8^{+2.4}_{-2.3}$&$ <4.4$&---&---&---&$<6.5 $&$4.7^{+2.3}_{-3.1}\tablenotemark{b} $\\
 \cline{1-10}
 $\rm Ne~X~Ly\beta$  &    10.239 & $2.8^{+1.2}_{-1.1}$&---&---&---&---&---&---&---\\ 
 $\rm Ne~X~Ly\alpha$  &   12.135 & $12.6^{+3.0}_{-2.6}$ &$21.8^{+7.8}_{-6.7}$& $27.2^{+12.1}_{-10.4}$ &---&---&---&$42.2^{+25.1}_{-19.6}$ &$30.1^{+13.5}_{-11.5}$\\ 
 $\rm Ne~IX~r$  &         13.447 & $<4.6 $                  &$1.6^{+5.8}_{-1.6}$&$ <10.5$&---&---&$32.5^{+64.9}_{-0.5} $&$<25.6 $&$8.2^{+15.5}_{-1.4}\tablenotemark{b} $\\
 $\rm Ne~IX~i$  &         13.551 & $29.8^{+5.4}_{-4.7}$ &$29.3^{+10.4}_{-8.8}$& $52.8^{+21.6}_{-14.3}$&---&---&$44.0^{+34.3}_{-25.5} $&$40.1^{+29.5}_{-22.0}$ &$21.3^{+13.8}_{-10.6} $\\
 $\rm Ne~IX~f$  &         13.699 & $2.5^{+2.7}_{-1.7} $ &$4.5^{+5.9}_{-4.3}$&$ <9.5$&---&---&$< 63.0$&$<35.8 $&$<20.6 $\\
 \cline{1-10}
 $\rm O~VIII~Ly\beta$&    16.005 & $2.6^{+2.8}_{-2.0}$&---&---&---&---&---&---&---\\
 $\rm O~VIII~Ly\alpha$&   18.968 & $42.1^{+13.6}_{-11.4}$& $52.8^{+29.0}_{-20.2}$&$31.8^{+32.8}_{-20.7}$ &---&---&$2.4^{+1.4}_{-1.1}$ &$107.0^{+90.6}_{-61.8}$& $82.1^{+51.6}_{-39.2}$\\
 $\rm O~VII~r$&           21.602 & $14.7^{+16.4}_{-13.1}$& $3.8^{+17.3}_{-3.7}$ &$ <26.8$&---&---&---&$<79.1 $&$<82.2 $\\
 $\rm O~VII~i$&           21.800 & $147.9^{+37.0}_{-31.8}$& $195.4^{+73.7}_{-60.0}$ &$430.8^{+150.2}_{-12.6}$&---&---&$<210.1 $&$338.7^{+247.1}_{-174.5} $&$149.1^{+107.4}_{-75.5} $\\
 $\rm O~VII~f$&           22.101 & $13.8^{+15.9}_{-11.2}$& $< 25.7$ &$ <20.1$&---&---&---&$<106.3 $&$<45.1 $\\
 \cline{1-10}
 $\rm N~VII~Ly\alpha$ &   24.781 & $88.7^{+29.3}_{-24.3}$& $52.1^{+43.6}_{-29.5}$&$79.5^{+90.5}_{-49.5}$&---&---&$124.7^{+205.8}_{-108.4}$&---&$78.7^{+95.4}_{-57.7}$\\
 $\rm N~VI~i$&            29.080 & $132.5^{+111.8}_{-112.4}$& $< 214.0$ &$ <684.5$ &---&---&---&---&---\\
\enddata
\label{tab-lines}
\tablenotetext{a}{Flux is in the units of $10^{-5} \rm ~photons~ s^{-1}~cm^{-2} $}
\tablenotetext{b}{67\% confidence level}
\end{deluxetable}
\end{landscape}

\begin{deluxetable}{l|cccc}
\footnotesize
\tablecolumns{5}
\tablecaption{He-like lines\tablenotemark{a}~ for \herx1}
\tablewidth{0pt}
\tablehead{
Ion                    &$\rm Si~XIII$ & $\rm Mg~XI$    & $\rm Ne~IX $    &$\rm O~VII $ \\
}
\startdata
&   \multicolumn{4}{c}{ID 2749}\\
\cline{1-5}
$G=\frac{(f+i)}{r}$    &  $4.3^{+6.6}_{-1.9} $ & $4.1^{+3.9}_{-1.6} $ & $> 7.1 $                 & $13.4^{+37.0}_{-7.4}$\\
$R=\frac{f}{i}$        &  $3.9^{+7.7}_{-1.8} $ & $0.5^{+0.3}_{-0.2} $ & $0.09^{+0.09}_{-0.07} $ & $0.6^{+0.1}_{-0.1} $\\
Cash dof               &  36.1/37                    & 17.4/16                    & 11.2/15 & 10.4/17\\
\cline{1-5}
&   \multicolumn{4}{c}{ID 3821}\\
\cline{1-5}
$G=\frac{(f+i)}{r}$    &---   & $4.2^{+119.8}_{-2.5} $ & $>3.9 $ & $>2.5$\\
$R=\frac{f}{i}$        &---   & $0.7^{+0.9}_{-0.5} $ & $<0.4 $ & $<0.5$\\
Cash dof               &---   & 18.3/23                    & 48.0/17   & 51.1/18\\
\cline{1-5}
&   \multicolumn{4}{c}{ID 3822}\\
\cline{1-5}
$G=\frac{(f+i)}{r}$    & $>3.1 $   & $>0.5 $  & $>2.7 $ & $>2.4$\\
$R=\frac{f}{i}$        & $>1.8 $   & $<1.4 $  & $<0.3 $ & $<0.3 $\\
Cash dof               &30.8/40      & 5.2/13     & 11.9/12  & 39.3/16\\
\cline{1-5}
&   \multicolumn{4}{c}{ID 4585-L}\\
$G=\frac{(f+i)}{r}$    &---   &---   &  $2.9^{+8.4}_{-1.6}$&--- \\
$R=\frac{f}{i}$        &---   &---   & $<1.3 $ &--- \\
Cash dof               &---   &---   & 37.1/22  &--- \\
\cline{1-5}
&   \multicolumn{4}{c}{ID 6149-L}\\
$G=\frac{(f+i)}{r}$    & $>0.1 $   & $>0.7 $  & $>4.2 $ & $3.8^{+3.9}_{-2.6}$\\
$R=\frac{f}{i}$        & $>0.2 $   & $<1.3 $  & $<0.9 $ & $0.3^{+0.4}_{-0.2}\tablenotemark{b} $\\
Cash dof               &39.3/41      & 15.9/21     & 14.5/19  & 14.1/11\\
\cline{1-5}
&   \multicolumn{4}{c}{ID 6150}\\
\cline{1-5}
$G=\frac{(f+i)}{r}$    & $>1.0 $   & $>1.8 $  & $>1.6 $ & $5.0^{+31.8}_{-3.0}$\tablenotemark{b}\\
$R=\frac{f}{i}$        & $1.5^{+3.6}_{-0.8}$\tablenotemark{b}   & $0.7^{+1.7}_{-0.6} $  & $0.3^{+0.5}_{-0.3} $\tablenotemark{b} & $<0.5 $\\
Cash dof               &41.7/31      & 1.9/12     & 27.4/26  & 11.3/10\\
\enddata
\tablenotetext{a}{The line center is fixed at the theoretical value and the line width is fixed as 0.001\AA.}
\tablenotetext{b}{at 67\% confidence level.}
\label{tab-He}
\end{deluxetable}

\newpage

%
%

%

%
%

\end{document}